\documentclass[onecolumn,11pt,draftcls]{IEEEtran} 
\usepackage{color}
\usepackage{amsbsy}
\usepackage{floatflt} 
\usepackage{url}
\usepackage{arydshln}
\usepackage{amsmath}
\usepackage{amssymb}
\usepackage{stackrel}
\usepackage{times}
\usepackage{graphicx}
\usepackage{xspace}
\usepackage{paralist} 
\usepackage{setspace} 
\usepackage{xypic}
\xyoption{curve}
\usepackage{latexsym}
\usepackage{theorem}
\usepackage{ifthen}
\usepackage{subfigure}
\usepackage[ruled,vlined]{algorithm2e}
\usepackage{booktabs}
\usepackage{algorithmic}\lightrulewidth=0.03em

\newcommand{\mypar}[1]{\vspace{0.03in}\noindent{\bf #1.}}

{\theoremheaderfont{\it} \theorembodyfont{\rmfamily}
\newtheorem{theorem}{Theorem}
\newtheorem{lemma}[theorem]{Lemma}

}

\def\O{{\mathcal O}}
\def\bfone{\bf 1}
\def\bfzero{{\bf 0}}

\title{Sensor Selection for Event Detection in Wireless Sensor Networks}
\author{Dragana Bajovi\'{c}, Bruno Sinopoli and Jo\~{a}o Xavier 
\thanks{Partially supported by grants SIPM PTDC/EEA-ACR/73749/2006 and
SFRH/BD/33517/2008 (through the Carnegie Mellon/Portugal Program
managed by ICTI) from Funda\c{c}\~{a}o para a Ci\^encia e Tecnologia and also by
ISR/IST plurianual funding (POSC program, FEDER).}
\thanks{D. Bajovi\'{c} is with the Institute for Systems and Robotics (ISR), Instituto Superior
T\'{e}cnico (IST), Lisbon, Portugal, and with the Department of Electrical and Computer Engineering,
Carnegie Mellon University, Pittsburgh, PA, USA {\tt\small dragana@isr.ist.utl.pt, dbajovic@andrew.cmu.edu}}%
\thanks{B. Sinopoli is with the Department of Electrical and Computer Engineering, Carnegie Mellon University,
Pittsburgh, PA, USA {\tt\small brunos@ece.cmu.edu}}%
\thanks{J. Xavier is with the Institute for Systems and Robotics (ISR), Instituto Superior T\'{e}cnico (IST), Lisbon, Portugal {\tt\small jxavier@isr.ist.utl.pt}}}

\begin{document}
\maketitle

\begin{abstract}
We consider the problem of sensor selection for event detection in wireless sensor networks (WSNs).
We want to choose a subset of $p$ out of $n$ sensors that yields the best
 detection performance. As the sensor selection optimality criteria, we propose the Kullback-Leibler and Chernoff distances
  between the distributions of the selected measurements under the two hypothesis. We formulate the maxmin robust sensor selection problem to cope with the uncertainties in distribution means.
   We prove that the sensor selection problem is NP hard, for both Kullback-Leibler and Chernoff criteria.
   To (sub)optimally solve the sensor selection problem, we propose an algorithm of affordable complexity.
    Extensive numerical simulations on moderate size problem instances (when
    the optimum by exhaustive search is feasible to compute) demonstrate the algorithm's near optimality in
    a very large portion of problem instances. For larger problems, extensive simulations demonstrate that
     our algorithm outperforms random searches, once an upper bound on computational time is set. We corroborate
     numerically the validity of the Kullback-Leibler and Chernoff sensor selection criteria, by showing that they lead to
     sensor selections nearly optimal both in the Neyman-Pearson and Bayes sense.
\end{abstract}
\hspace{.43cm}\textbf{Keywords:} sensor selection, event detection, robust optimization, Chernoff distance, Kullback-Leibler distance

\newpage
\section{Introduction}
\label{sec-Introduction}
Wireless sensor networks (WSNs) operate with limited power and communication resources.
 When observing phenomena with WSNs, a major challenge is to balance
  the tradeoff between the quality and the cost of operation.
   A fundamental problem of this kind, the sensor selection problem, is how to optimally select a limited subset of sensors (hence limiting the operation cost) that gives the most valuable information about the observed phenomena.

\mypar{Problem statement} This paper studies the sensor selection problem for event detection in WSNs. Nature can be in one of two states: $H_1$ (event occurring, e.g., target present) and $H_0$ (event not occurring, e.g., target absent).
 A WSN, composed of $n$ sensors, instruments the nature. The distribution of
 the $n$-dimensional measurement vector is assumed Gaussian under the two hypothesis,
  with different means $m_i$ and covariances $S_i$, $i=0,1$. We assume that, due to inherent WSN
   constraints, such as power, only $p$ (out of $n$) sensors can sense and transmit their readings to a fusion node; based on the received $p$ readings, the fusion node performs detection, i.e., decides which of the two hypothesis is true. We ask the following question:
 \emph{Which $p$~sensors should be chosen to achieve the best detection performance?}

Each possible $p$-sensor selection induces a $p$ dimensional Gaussian distribution $\pi_i$ (of selected sensors)
under hypothesis $H_i$, $i=0,1$. Intuitively, $p$-sensor selection that yields more distant distributions $\pi_1$ and $\pi_0$
  leads to better detection. Hence, we propose, as sensor selection optimality criteria: 1) the Kullback-Leibler (KL)
  distance; and 2) the Chernoff (C) distance between $\pi_1$ and $\pi_0$. In practice, the distribution parameters
  $(m_i, S_i)$ are estimated from training data, and may not be known exactly, but only within an uncertainty region.
We thus formulate the robust maxmin sensor selection problem of maximizing the KL (or C) distance between
the selected distributions $\pi_1$ and $\pi_0$, for the worst case of parameter
drifts. In this paper, we address the case when only the means of the two distributions are uncertain, relegating the general case for
future work.

\mypar{Contributions} The problem of evaluating the best $p$-sensor subset is combinatorial; checking
 over all $n \choose p$ possible combinations becomes infeasible when
  $n$ and $p$ are sufficiently large. We indeed prove that the KL and C sensor selection
  problems are NP hard; hence, it is unlikely to find an efficient algorithm that solves
  large instances of these problems. To (suboptimally in general) solve the sensor selection problems,
  we propose a computationally affordable algorithm. For example, to select 10 out of 100 sensors,
   our algorithm takes only few seconds on a current generation personal computer. There is no theoretical guarantee that our
   algorithm produces an optimal or near optimal solution; however,
    extensive numerical experiments demonstrate
    that our algorithm produces an optimal (or near optimal) solution in most cases.

The KL and C distances are only heuristic optimality measures
for the detection performance; the exact (yet in our case intractable)
 criterion is the probability of detection error (either in Neyman-Pearson
or Bayes sense.) However, interestingly
enough, we show by simulations that optimizing the KL and C
 distances yields to sensor selections that are very close
 to the optimal probability of error (either in Bayes or Neyman-Pearson sense).
 Moreover, they are often indeed optimal.

Selecting $p$ out of $n$ sensors is equivalent to finding the
$n \times p$ selection matrix with 0/1 entries that maps
the $n$-dimensional vector $x$ of all measurements to the $p$-dimensional vector $y$ of selected measurements.
Our methodology for solving the combinatorial
sensor selection problems relies on enlarging the search space from the 0/1 selection matrices to
the set of Stiefel matrices (the matrices with orthonormal columns). Then, after solving for the Stiefel matrix,
we project it back to the set of 0/1 selection matrices.
 The relaxed Stiefel problem corresponds to finding the
linear map ${\mathbb R}^n \rightarrow {\mathbb R}^p$ which maximizes the
(worst-case) KL (C) distance between the projected distributions in the lower $p$-dimensional space.
To our best knowledge, existing work on this topic either
does not solve the problem in full generality (i.e., unequal
means and covariance matrices) or does not guarantee global
optimality of their solutions (see~\cite{RandHerr,Scharf} for
problems involving Chernoff distance and the closely related J-divergence). A major contribution of this paper is that we solve this nonconvex problem globally for the
case $p = 1$, and in full generality,
 by reducing it to scalar (1D) problem over a compact interval.
We tackle the generic case $p > 1$ via an incremental, greedy approach, based on the 1D case, which provides
near optimal result with small computational cost.

This paper is related to our prior work~\cite{CDC-paper, allerton-dasica},
which also considers sensor selection for event detection, but only based on the KL distance.
 With respect to KL distance, this paper provides a new heuristic with reduced
 complexity; more importantly, this paper studies the problem with respect to the Chernoff distance, which we
 did not consider in~\cite{CDC-paper, allerton-dasica}.
 With respect to~\cite{CDC-paper, allerton-dasica}, this paper also contributes by validating the KL and C distances as good optimality criteria by showing their near optimality in the error probability sense,
  and by establishing NP hardness of the corresponding sensor selection optimization problems.

Finally, we would like to note that the KL-based and C-based sensor selection problems could be, in principle, globally solved (inefficiently) by, e.g., branch and bound methods~\cite{MaxEntropySampling,branchandbound}.
   However, the computational time of such methods is often very long, even for modest values of
   $n$ and $p$. We discuss in more detail the challenges to solve the sensor selection problems
   that we address at the end of subsection~{II}-B.

\mypar{Review of existing work on sensor selection}
Sensor selection problems have been extensively studied in different contexts,
including target tracking (e.g.,~\cite{Zhao}),
target localization (e.g.,~\cite{Kaplan}), robotics (e.g.,~\cite{Robotics-Hovland}),
and wireless sensor networks (e.g.~\cite{Buczak, Chu}). Generally, all their work aims to
optimize certain measure of performance of the system
(e.g., utility~\cite{utility}, information theoretic measure~\cite{Zhao,Ertin}, estimation error~\cite{Boyd}) subject
to energy constraints (e.g., limited number of sensors to be selected~\cite{Boyd}).
Sensor selection can be geometric-based (e.g.,~\cite{Isler}), or information-theoretic
based (e.g.,~\cite{Zhao,Ertin}). Our work belongs to the class of sensor selection problems for inference tasks, i.e., sensor selection for estimation and detection, and it adopts the information-theoretic point of view. Reference~\cite{Boyd} considers the problem of sensor selection for parameter estimation in WSNs,
 proposing to select the subset of $p$ (out of $n$) sensors that minimize the determinant
of the estimator covariance matrix. Reference~\cite{DistributedSenSel}
 proposes distributed algorithms to (suboptimally) solve the sensor selection
 problem formulated in~\cite{Boyd}. Reference~\cite{Giannakis} addresses the problem of selecting
 the maximal number of reliable sensors
 for estimation.
Reference~\cite{Sastry} shows, through the optimal experiment design framework~(\cite{Fedorov}) and using convex analysis, that optimal estimation is achievable by using only a relatively small number of sensors.

\mypar{Paper organization} Section~{II} introduces the model and formulates the
sensor selection optimization problems. Section~{III} details the
algorithms for solving the robust sensor selection problems, in the presence of uncertainties in the means of the distributions.
 Section~{IV} considers the special, yet important case, when there are no uncertainties
 in the distribution parameters. Section~{V} demonstrates
 numerically that the KL and the C distances are appropriate metrics for sensor selection. Section~{VI} shows near optimal performance
  of the proposed algorithms. Finally, Section~{VII} concludes the paper.
\vspace{-0.2em}
\section{Problem model: Formulation of the sensor selection optimization problem}
\label{sec-Problem-formulation}
\subsection{Problem model}
We assume that nature can be in one of two states: $H_1$--event occurring, and $H_0$--event not occurring.
Let $x \in {\mathbb R}^n$ denote the vector that collects all sensor measurements (one measurement per sensor).
We assume that $x$ is Gaussian under the hypothesis $H_1$ (respectively, $H_0$,)
with generic means and covariances $\left( m_1, S_1\right)$ (respectively, $(m_0, S_0)$), i.e.,
\vspace{-0.7em}
\[
\begin{array}{ccc} H_0 & : & x \sim {\mathcal N}(m_0,S_0) \\ H_1 & : & x \sim {\mathcal N}(m_1,S_1) \end{array},
\]
where $\mathcal N (\mu, \Sigma)$ denotes Gaussian distribution with the mean vector $\mu$ and the covariance matrix $\Sigma$. The Gaussian assumption on $x$ is standard and can be, in many applications, justified, e.g, by central limit theorem arguments, see, e.g.,~\cite{sayed_sdp, sayed-CRs}. Noise correlation (i.e., non diagonal covariance matrices) is important to take into account in dense deployments of WSNs. We note that our formulation allows for different covariances under two hypothesis ($S_1 \neq S_0$); in many applications, e.g., power-based detection of primary users for cognitive radios (see~\cite{sayed-CRs} for details), accounting for different covariances is essential.\footnote{Generally, our model applies also if $x$ contains samples from multiple sensors over \emph{ multiple sample times}, that is, several entries in $x$ correspond to same sensor. In such scenario vector $x$ accounts both for temporal and spatial
correlation between observations. To focus the presentation, however, we assume that
 different entries of $x$ correspond to different sensors.}

\mypar{Sensor selection} Sensors transmit their measurements to a fusion node,
which conducts the hypothesis test.
 Due to power constraints, only $p$~sensors, $p<n$, perform their measurements
and transmit them to the fusion node. We address the problem of selecting the $p$
sensors that guarantee the best detection performance.
Mathematically, selecting $p$ out of $n$ sensors can be represented by a linear map
${\mathbb R}^n \rightarrow {\mathbb R}^p,\,\,x \mapsto y = E^\top x$, where $E \in {\mathbb R}^{n \times p}$ is a rank-$p$ matrix that has exactly one unit entry per column, corresponding to a chosen sensor, and the other entries in columns being zero. The columns of $E$
 are orthonormal, i.e., $E^\top E=I_p$, where $I_p$ denotes the $p\times p$ identity matrix. We refer to matrix $E$ as the sensor selection matrix.

\mypar{Hypothesis test induced by $E$}
Conditioned on $H_i$, $i=0,1$, $y$ is a linear transformation of a Gaussian vector $x$. Thus, $y$,
under $H_i$, has the following distributions:
\vspace{-0.4em}
\begin{equation}
\label{eq-projected-Gaussian-distributions}
H_0 \;\;\; : \;\;\; y \sim {\mathcal N}( E^\top m_0, E^\top S_0 E ) \\
\end{equation}\vspace{-2.7em}
\begin{equation*}
H_1 \;\;\; : \;\;\; y \sim {\mathcal N}( E^\top m_1, E^\top S_1 E). \nonumber
\end{equation*}
The fusion node performs the following log-likelihood ratio (LLR) decision test:
\begin{equation}
\label{eq-induced-LLR}
\log  \frac{f_1(y;E)}{f_0(y;E)} \stackrel[H_0]{H_1}{\gtrless}\gamma,
\end{equation}
where $f_i(\cdot \,; E)$, $i=0,1$, is the density of $\mathcal{N} \left( E^\top m_i, E^\top S_i E \right)$ and $\gamma \in {\mathbb R}$ is the test threshold,~\cite{Scharf}.
\vspace{-0.5em}
\subsection{Formulation of the sensor selection optimization problem}
\mypar{Sensor selection optimality criteria}
Detection performance is, as noted above, generally quantified by the error probabilities, $P_{\mathrm{FA}}$, $P_{\mathrm{D}}$ and ($P_{\mathrm{e}}$). However, in the problem that we consider, none of the probabilities above admit closed
 form expression, and their minimization with respect to sensor selection is a very hard problem. As sensor selection optimality criteria, we propose the Kullback-Leibler (KL) and the Chernoff (C) distance between the tested distributions. Given two distributions, with densities $f_1$ and $f_0$, KL and C distances measure dissimilarity between $f_1$ and $f_0$, and they are defined as follows:
\begin{eqnarray*}
D_{\mathrm{KL}}\left(f_1\,\|\,f_0\right)&:=&\int \log \frac{f_1(x)}{f_0(x)}f_1(x)dx \\
D_{\mathrm{C}}\left(f_1, \,f_0\right)&:=&\max_{s\in[0,1]}  -\log \int f_1^s(x) f_0^{1-s}(x) dx.
\end{eqnarray*}
Thus, our goal is to find a sensor selection that yields the maximal KL or C distance between the projected Gaussian distributions, ${\mathcal N}(E^\top m_1, E^\top S_1 E)$ and ${\mathcal N}(E^\top m_0, E^\top S_0 E)$ (see~\eqref{eq-projected-Gaussian-distributions}).

We are motivated to choose KL and C distances, as sensor selection optimality criteria, by two fundamental results from detection theory: Chernoff-Stein lemma and Chernoff lemma.
Chernoff-Stein lemma (resp. Chernoff lemma) states that, when the number of independent identically distributed (i.i.d.) observations grows,
the rate of exponential decay of probability of false alarm, $P_{\mathrm{FA}}$, (resp. probability of error, $P_{\mathrm{e}}$,) of the Neyman-Pearson optimal (resp. Bayes optimal) test equals the KL (resp. C) between the two distributions.
Thus, for large number of samples,
more distant distributions (in either KL or C sense) lead to better detection performance.
 Probabilistic distance measures have been often used in the literature as heuristics for detection problems
 (see, e.g.,~\cite{RandHerr,ComparisonChernoff} for applications in linear dimensionality reduction) and have
 shown excellent results, even when the number of samples is very small or even equal to one, see~\cite{ComparisonChernoff}.
 In section~{V}, we demonstrate by numerical tests that the KL and C distances are indeed
 excellent criteria for sensor selection, exhibiting near optimal performance in the probability of error sense.
  Section~{V} shows that, generally, C distance has an advantage in the regimes of high probability of detection ($P_{\mathrm{D}}$) (upper part of the receiver operating characteristics (ROC) curve), while KL has an edge in the regimes of low $P_{\mathrm{FA}}$ (lower left part of the ROC curve).

\mypar{Robustness against uncertainty in distribution means}
We consider the case when the true distribution parameters $(m_i,S_i)$, $i=0,1$ are not exactly known at the fusion node detector~\eqref{eq-induced-LLR}. Fusion node has their estimates, $\left(\widehat m_i, \widehat S_i \right)$, which can be obtained, e.g., in the network training phase. Thus, there is a mismatch between the distribution used by the fusion node detector and the distribution that generates the observations. Our goal is to design a sensor selection that yields detection~\eqref{eq-induced-LLR} robust to these mismatches. In this paper, we restrict our attention to the case where only the mean values are uncertain (the true covariance matrices are known), and we allow the mean values to drift in the following ellipsoidal uncertainty regions:
\begin{equation}
m_i \in \mathcal{E} \left( \widehat m_i, k_i\, S_i^{-1} \right)\,\,\,i=0,1.
\label{eq-conf-ellipsoids}
\end{equation}
Here $\widehat m_i$ denotes the estimated mean vector, $i=0,1$, $\mathcal{E}(a,A)$ ($A$ is a positive definite matrix) denotes the ellipsoid
\vspace{-0.5em}
\[
\mathcal{E}\left( a,A \right) = \{x \in {\mathbb R}^n: \,\, (x-a)^\top A (x-a) \leq 1\}.
\]
and the parameter $k_i \in (0, +\infty]$ is a free parameter which controls the ``size'' of the uncertainty region, e.g, if $k_i = +\infty$,
there is no uncertainty: $\widehat m_i= m_i$. The orientations of the uncertainty ellipsoids in~\eqref{eq-conf-ellipsoids} are induced by the covariance matrices $S_0$ and $S_1$. This choice of the form of uncertainty regions is motivated by the following fact: if the means are estimated via the sample mean estimator based on $N$ i.i.d. observations (the minimum variance unbiased estimate for Gaussian distributions),
then the covariance of the estimate $\widehat m_i$ equals $\frac{1}{N} S_i$, for $i=0,1$. A standard measure of the uncertainty, the confidence region~(\cite{Scharf}), is for $m_i$ given by~\eqref{eq-conf-ellipsoids}.
The scaling constants $k_0$ and $k_1$ are, in this context, proportional to $N$; also, if $N$ is known, $k_0$ and $k_1$ can be used to design the uncertainty regions of desired confidence levels.

We address the uncertainties in the mean vectors adopting the worst case approach. That is, we search for the sensor selection that gives the maximal distance (KL and C) for the worst case of the mean parameter drift.

\mypar{Optimization problems}
\label{sec-Optimization-problems}
We introduce the following two functions, $f_\mathrm{KL}$ and $f_\mathrm{C}$, that capture the dependence of the KL and the C distance on the selection matrix $E$ and the mean vectors $m_0$ and $m_1$:
\begin{eqnarray*}
f_\mathrm{KL}(E, m_0, m_1)&:=&D_\mathrm{KL}\left( {\mathcal N}(E^\top m_1, E^\top S_1 E) \|  {\mathcal N}(E^\top m_0, E^\top S_0 E) \right)\\
\mbox{max}_{s\in[0,1]} f_\mathrm{C}(s, E, m_0, m_1)&:=&D_\mathrm{C}\left( {\mathcal N}(E^\top m_1, E^\top S_1 E) ,\,  {\mathcal N}(E^\top m_0, E^\top S_0 E) \right).
\end{eqnarray*}
The robust sensor selection optimization problems are then given as follows:
\begin{align}
& \begin{array}[+]{ll}
\mbox{maximize} & \mbox{min}_{m_0 \in \mathcal{E} \left( \widehat m_0, k_0\, S_0^{-1} \right), m_1 \in \mathcal{E} \left( \widehat m_1, k_1\, S_1^{-1} \right)} f_\mathrm{KL}(E, m_0, m_1) \\
\mbox{subject to} & E_{ij} \in \{0,1\} \\  & E^\top E = I_{p}
\end{array}\label{op-KL}\\
& \begin{array}[+]{ll}
 \mbox{maximize} & \mbox{min}_{m_0 \in \mathcal{E} \left( \widehat m_0, k_0\, S_0^{-1} \right), m_1 \in \mathcal{E} \left( \widehat m_1, k_1\, S_1^{-1} \right)} \,\, \mbox{max}_{s\in[0,1]} f_\mathrm{C}(s, E, m_0, m_1)\\
\mbox{subject to} & E_{ij} \in \{0,1\} \\  & E^\top E = I_{p}
\end{array}.\label{op-C}
\end{align}
\vspace{-10pt}
It can be shown that ($\mathrm{tr}(\cdot)$ and $\left|\cdot\right|$ denote the trace and the determinant, respectively):
\begin{align}
\label{KL-cost-fcn}
f_\mathrm{KL}(E, m_0, m_1) =
\frac{1}{2}\, \left\{  (m_1-m_0)^\top E {\left(E^\top S_0 E\right)}^{-1} E^\top (m_1 - m_0)
+   \mathrm{tr}\left(  {\left(E^\top S_0 E\right)}^{-1} E^\top S_1 E  \right)\;\;
\right. \nonumber \\
\left. - \log\frac{ \left| E^\top S_1 E \right|}{ \left| E^\top S_0 E \right|} - p \right\}\\
\label{C-cost-fcn}
f_\mathrm{C}(s, E, m_0, m_1) = \frac{1}{2}\,\left\{ s(1-s)(m_1-m_0)^\top E \left(sE^\top S_0 E +(1-s) E^\top S_1 E \right)^{-1} E^\top(m_1-m_0)  \right. \;\;\;\, \nonumber \\
\left. -\log \frac {\left|E^\top S_0 E\right|^s \left|E^\top S_1 E\right|^{1-s}}{\left|sE^\top S_0 E +(1-s) E^\top S_1 E\right|}\right\}.
\end{align}

Optimization problems~\eqref{op-KL}~and~\eqref{op-C} are combinatorial. When $n$ and $p$ are small, a simple method for solving them is exhaustive search that checks all $n \choose p$ sensor subsets (i.e. selection matrices). For large $n$ and $p$, however, this method becomes computationally infeasible.
 Indeed, we have the following result, which we prove in the Appendix~\ref{App-NP-hardness}:
 \vspace{-0.5em}
 \begin{theorem}
 \label{theorem-NP-hardness}
 The optimization problems~\eqref{op-KL}~and~\eqref{op-C} are NP hard (even when $k_0=k_1=+\infty$ and $S_1=S_0$).
 \end{theorem}
  \vspace{-0.3em}
In principle, besides exhaustive search, problems~\eqref{op-KL}~and~\eqref{op-C} can be (inefficiently) solved by branch and bound methods, see, e.g.,~\cite{branchandbound}. However, complexity of these methods relies strongly on the choice of bounds on the cost function;  finding tight bounds is a hard problem itself.
An interesting method for solving a problem somewhat similar to ours is proposed in~\cite{Guestrin}. In~\cite{Guestrin}, the authors
address the problem of finding the most informative locations, for future sensor placements, in a discretized Gaussian field; the measure of informativeness that the authors propose is the mutual information between sensed and un-sensed locations. The authors show that the mutual information is a submodular function, which assures that simple strategy of choosing sensor by sensor (greedy) gives a solution within $1-\frac{1}{e}$ of
the optimum. In our problems, however, such a bound on the greedy strategy (of selecting sensors one by one) does not hold, as our cost functions (KL and C distance) are not submodular (see the proof in Appendix~\ref{App-Not-submodular-CounterEx}). In fact, in our problems, greedy performs poorly in many cases. The reason for this lies in the fact that the correlations--an information
that greedy discards can play an important role in discerning between the two hypothesis.
\vspace{-0.5em}
\section{Sensor selection algorithms}
\label{sec-Algorithms}
In this section, we present our algorithms, R--KL (robust KL based selection) and R--C (robust C-based selection), that, respectively, solve the problems~\eqref{op-KL} and~\eqref{op-C}.
First, we explain the methodology and the structure of the algorithms, and we explain the geometrical intuition behind this methodology. Subsections~\ref{subsect-R-KL} and~\ref{subsec-R-C} detail the R--KL and R--C algorithms.

\mypar{Algorithms methodology and structure}
Geometrically, one combination of $p$ sensors defines one $p$-dimensional subspace of ${\mathbb{R}}^n$, spanned by a set of canonical basis vectors corresponding to the chosen sensors. We call this subspace a canonical subspace. The cost functions~\eqref{KL-cost-fcn} and~\eqref{C-cost-fcn} depend on $E$ only through its range, and problems~\eqref{op-KL} and~\eqref{op-C}, in this sense, search for the best subspace, among $n \choose p$ canonical subspaces, on which the original distributions should be projected. We relax these combinatorial problems by allowing for projections to arbitrary $p$-dimensional subspaces. Mathematically, this translates into replacing the set of 0/1 selection matrices with the set of $n\times p$ Stiefel matrices (that represent all $p$-dimensional subspaces). Then, we use a solution of the relaxed problem and ``round'' it by the closest canonical subspace.

We call the first phase of our algorithm, that solves the relaxed Stiefel problem, the \emph{Relaxation phase}; the second phase, in which we find the closest canonical subspace, is the \emph{Projection phase}. Finally, the last, third step in our algorithms, the \emph{Refinement phase}, refines the solution by performing local optimization.
\vspace{-1em}
\subsection{Algorithm for the robust Kullback-Leibler based selection: the R--KL algorithm}
\label{subsect-R-KL}
\subsubsection{Relaxation phase}
\label{subsubs-relax-phase}
We solve the following Stiefel relaxation of the problem~\eqref{op-KL}:
\begin{equation}
\begin{array}[+]{ll}
\mbox{maximize} & \min_{m_0 \in \mathcal{E} \left( \widehat m_0, k_0\, S_0^{-1} \right), m_1 \in \mathcal{E} \left( \widehat m_1, k_1\, S_1^{-1} \right)}f_\mathrm{KL}(E, m_0, m_1) \\
\mbox{subject to} & E^\top E = I_{p}.
\end{array}\label{eq-KL-op-relaxation}
\end{equation}
The relaxed problem~\eqref{eq-KL-op-relaxation} is nonconvex and still difficult to solve. We globally solve this problem for the case $p=1$, i.e., we find the best 1-dimensional projection. General, $p>1$ case, is addressed by a greedy approach, using our $1-D$ tool.
We first detail the algorithm that solves $p=1$ case.

\emph{(1.1) Case $p=1$: global solution}
For $p=1$, the constraint set of Stiefel matrices reduces to a sphere in $\mathbb R^n$ and the problem~\eqref{eq-KL-op-relaxation} takes a simplified form:
\begin{equation}
\begin{array}[+]{ll}
\mbox{maximize} & \mbox{min}_{\substack{m_0 \in \mathcal{E} \left( \widehat m_0, k_0\, S_0^{-1} \right),\\ m_1 \in \mathcal{E} \left( \widehat m_1, k_1\, S_1^{-1} \right)}}\frac{1}{2}\, \left\{\frac{\left(e^\top (m_1 - m_0)\right)^2}{e^\top S_0 e}+
\frac{e^\top S_1 e}{e^\top S_0 e}-
\log \frac { e^\top S_1 e}{e^\top S_0 e} - 1 \right\} \\
\mbox{subject to} & e^\top e = 1
\end{array}.\label{op-KL-relaxation-p-eq-1}
\end{equation}
Our major contribution is showing that the problem~\eqref{op-KL-relaxation-p-eq-1} reduces to a search over a compact (one-dimensional) interval. We achieve this by a series of judicious problem reformulations, and by invoking convexity
of quadratic mappings, \cite{Brickman,Polyak}. It is important to note that the original problem~\eqref{op-KL-relaxation-p-eq-1}
has in general very high dimensionality (equal to the total number of sensor in the network $n$); also, due to its nonconvexity, it is very difficult to solve globally. By doing reformulations, we manage to map it to a tractable, scalar problem.
The next Lemma, proved in~\cite{allerton-dasica}, states the first step towards this goal. It shows that a solution of~\eqref{op-KL-relaxation-p-eq-1} can be reconstructed after solving a 2 dimensional problem~\eqref{op-psi-KL}.
\vspace{-0.5em}
\begin{lemma}
\label{lemma-first-step-KL}
Suppose $(x^\star,\,y^\star)$ solves
\begin{equation}
\begin{array}[+]{ll}
\mbox{maximize} & \psi_{\mathrm{KL}}(x,y) \\
\mbox{subject to} & (x,y) \in \mathcal{R}
\end{array} \label{op-psi-KL}
\end{equation}
where
\vspace{-1em}
\begin{equation}
\label{eqn-psi-KL}
\psi_{\mathrm{KL}}(x,y) = x - \log x +
\left\{ \left(
 \sqrt{y} - \frac{1}{\sqrt{k_1}}  \sqrt{ x } - \frac{1}{\sqrt{k_0}} \right)^{+} \right\}^2,\\
 \end{equation}
\begin{equation}
 \mathcal{R} = \left\{ (x,y) \in {\mathbb R}^2:\,\,  x = v^\top S v,\,\,y=v^\top M  v,  \mathrm{\,\,\,\,for\,\,some\,\,} v \in {\mathbb R}^n,\,\, v^\top v =1 \right \}, \label{def-R}
\end{equation}
$x^+=\max\left(0,x\right)$, $S=S_0^{-1/2}S_1 S_0^{-1/2}$, $m=S_0^{-1/2} (m_1-m_0)$, $M=m m^\top$. Let $v^\star \in {\mathbb R}^n$ be an unit-norm vector that generates $x^\star$ and $y^\star$, i.e. $x^\star={v^\star}^\top S v^\star$ and $y^\star={v^\star}^\top M v^\star$.
Then, $e^\star := S_0^{-1/2}v^\star/\|S_0^{-1/2}v^\star\|$ solves~\eqref{op-KL-relaxation-p-eq-1}.
\end{lemma}
Lemma~\ref{lemma-first-step-KL} says that, in order to solve~\eqref{op-KL-relaxation-p-eq-1}, it suffices to search over the set $\mathcal R  \subset {\mathbb R}^2$. For $n \geq 3$, the set $\mathcal{R}$ is compact and convex, as the image set of a unit sphere under two quadratic mappings, see~\cite{Brickman}. Note that, since $\psi_{\mathrm{KL}}(\cdot)$ is continuous and ${\mathcal R}$ is compact, there is a global maximizer, by the Weierstrass theorem. Next lemma further simplifies the search by asserting that the boundary of~${\mathcal R}$ contains a global maximizer. For the proof of Lemma~\ref{lemma-max-at-boundary}, see~\cite{allerton-dasica}.
\vspace{-0.5em}
\begin{lemma}
\label{lemma-max-at-boundary}
The boundary $\partial \mathcal{R}$ of the set $\mathcal{R}$ contains a global maximizer of~\eqref{op-psi-KL}.
\end{lemma}
The boundary of $\mathcal{R}$ is a closed curve in ${\mathbb R}^2$. Our strategy consists in circulating along~$\partial R$ to spot a global maximizer. More precisely, we will sample~$\partial R$ with a finite set of points and pick the best point. To implement this strategy, we borrow the following theorem from~\cite{Polyak}.
\vspace{-0.5em}
\begin{theorem}[\cite{Polyak}]
\label{theorem-polyak}
Let $n \geq 3$ and let~$A, B$ be $n \times n$ symmetric matrices. Let
\vspace{-0.5em}
\begin{equation*}
\mathcal{R}(A,B) = \left\{ (x,y) \in {\mathbb R}^2:\,\,  x = v^\top A v,\,\,y=v^\top B v,
\mathrm{\,\,\,\,for\,\,some\,\,} v \in {\mathbb R}^n,\,\, v^\top v =1 \right \}.
\end{equation*}
\vspace{-0.3em}
For $t \in [0, 2 \pi]$, let $C(t) = A \cos t + B \sin t$ and let $\lambda_{\min}(t)$ be the minimal eigenvalue of the matrix $C(t)$ and $u_{\min}(t)$ an associated unit-norm eigenvector.
Suppose that $\lambda_{\min}(t)$ is a simple eigenvalue of $C(t)$ for all $t \in [0, 2\pi]$.
Then, the set $\mathcal{R}(A,B)$ is strictly convex and its boundary is given by
\vspace{-0.5em}
\begin{equation*}
 \partial \mathcal{R}(A,B)=  \left\{{(x(t),y(t)):\,\,t \in [0,2\pi ],\,\,\,\,x(t) = u_{\min}(t)^ \top  A u_{\min}(t),\,\,y(t) = u_{\min}(t)^ \top  B u_{\min}(t)} \right\}. \label{boundary}
\end{equation*}
\end{theorem}
\vspace{-1em}
\mypar{Parametrization of $\partial \mathcal{R}$}
Theorem~\ref{theorem-polyak} shows that the set $\mathcal R(A,B)$ (in our context, the set $\mathcal R=\mathcal R(S, mm^\top) $) can be parameterized by moving a single parameter $t$ over the compact interval $[0,2\pi]$. The theorem assumes that $\lambda_{\min}(t)$ is simple for all $t$, and, consequently, that $\mathcal R(A,B)$ is strictly convex. However, a parametrization of the boundary when this condition is not satisfied is readily available, as we explain next. Applied to our set~${\mathcal R}$ in~\eqref{def-R}, this leads to the following procedure:
\begin{enumerate}
\setlength{\itemsep}{1pt}
\item
generate the points
\begin{equation*}
(x_k, y_k) = (u_k^\top S u_k, u_k^\top  m m^\top u_k), \quad k = 1, 2, \ldots, K,
\end{equation*}
where $u_k$ denotes an unit-norm eigenvector corresponding to the minimal eigenvalue of
\begin{equation*}
C_k = S \cos \left( (k-1) 2 \pi / K \right) + m m^\top \sin \left( (k-1) 2 \pi / K \right).
\end{equation*}
Here, $K$ is the user-defined grid size and $\left\{ (x_k, y_k)\,:\,k = 1, \ldots, K \right\}$ is an initial sample of~$\partial {\mathcal R}$;
\item
if the distance between two consecutive points $(x_k, y_k)$ and $(x_{k+1}, y_{k+1})$ is greater than a prescribed threshold, interpolate the line segment which connects them, i.e., consider
\begin{eqnarray*}
\left( x_k^{(j)}, y_k^{(j)} \right) = ( 1 - j/J ) \left( x_k, y_k \right) + j/J  \left( x_{k+1}, y_{k+1} \right), \quad j = 0, 1, \ldots, J.
\end{eqnarray*}
\end{enumerate}
In summary, our sampling of~$\partial {\mathcal R}$ is $\widehat{\partial {\mathcal R}} = \left\{ (x_k, y_k) \right\} \cup \left\{ x_k^{(j)}, y_k^{(j)} \right\}.$

\mypar{Solving~\eqref{op-KL-relaxation-p-eq-1}}
 We explained how to solve~\eqref{op-psi-KL} by parameterizing $\partial {\mathcal R}$. Now we explain how a solution of~\eqref{op-psi-KL}, $x^\star, y^\star$, is sufficient to reconstruct $v^\star$, a solution of~\eqref{op-KL-relaxation-p-eq-1}. Vector $v^\star$ in Lemma~\ref{lemma-first-step-KL} can be found as follows. Let
 $$(x^\star, y^\star) \in \arg\max_{(x,y) \in \widehat{\partial {\mathcal R}}}\,\psi_{\mathrm{KL}}(x,y).$$ That is, $(x^\star, y^\star)$ denotes the best point in~$\widehat{\partial {\mathcal R}}$. If $(x^\star, y^\star) \in \left\{ (x_k, y_k) \right\}$, say $(x^\star, y^\star) = ( x_{k^\star}, y_{k^\star} )$, then we can take $v^\star$ as an unit-norm eigenvector associated with the minimal eigenvalue of $C_{k^\star}$. Otherwise, $(x^\star, y^\star) \in \left\{ (x_k^{(j)}, y_k^{(j)}) \right\}$, and we need to solve the system of~$3$ quadratic equations: \vspace{-0.5em}\begin{equation}v^\top S v = x^\star,\,\, v^\top m m^\top v=y^\star,\,\,v^\top v=1, \label{quadratics} \end{equation}\vspace{-0.35em}
with respect to (w.r.t.)~$v$. Any solution can be taken as~$v^\star$. It can be shown that~\eqref{quadratics} can be efficiently solved by solving a convex problem.

\emph{1.2) Case $p>1$: greedy algorithm}
Optimization problem~\eqref{eq-KL-op-relaxation} for the case $p > 1$ is very difficult to solve globally; we propose a greedy, suboptimal approach. We construct the columns of the matrix $E=\left[e_1\, e_2\, \ldots\,e_p\right]$ one by one (in the order $e_1,\, e_2, . . . $). We construct the $j$-th column by solving~\eqref{op-KL-relaxation-p-eq-1}, with the constraint that the column $e_j$ must be orthogonal to the previously determined columns $e_1,\,e_2,\,\ldots e_{j-1}$, i.e., we solve:
\begin{equation}
\begin{array}[+]{ll}
\mbox{maximize} & \min_{m_0 \in \mathcal{E}(\widehat {m_0}, k_0 S_0^{-1}),\,\, m_1 \in \mathcal{E}(\widehat {m_1} , k_1 S_1^{-1})}
f_{\mathrm{KL}}(e, m_0, m_1, S_0, S_1) \\
\mbox{subject to} & e^\top e = 1 \\ & e^\top e_i = 0, \quad i = 1, \ldots, j-1.
\end{array} \label{optim-problem-oneDi}
\end{equation}
Let $U^{(j)} \in {\mathbb R}^{n\times (n-j+1)}$ be a matrix with orthonormal columns which spans the orthogonal complement of~${\sf span}\left\{ e_1, \ldots, e_{j-1} \right\}$. The restrictions in~\eqref{optim-problem-oneDi} mean that $e = U^{(j)} e^{(j)}$ for some unit-norm $e^{(j)} \in {\mathbb R}^{n-j+1}$. This means that~\eqref{optim-problem-oneDi} corresponds to
\begin{equation}
\begin{array}[+]{ll}
\mbox{maximize} & \min_{m_0 \in \mathcal{E}(\widehat {m_0}, k_0 S_0^{-1}),\,\, m_1 \in \mathcal{E}(\widehat {m_1}, k_1 S_1^{-1})}
f_{\mathrm{KL}}( U^{(j)} e^{(j)}, m_0, m_1, S_0, S_1) \\
\mbox{subject to} & {e^{(j)}}^\top e^{(j)} = 1 .
\end{array} \label{optim-problem-oneDi2}
\end{equation}
The problem~\eqref{optim-problem-oneDi2} is equivalent to~\eqref{optim-problem-oneDi3} (see~\cite{allerton-dasica})
\begin{equation}
\begin{array}[+]{ll}
\mbox{maximize} & \min_{ m_0^{(j)} \in \mathcal{E}\left(\widehat {m_0^{(j)}}, k_0 \left( S_0^{(j)} \right)^{-1} \right),\,\, m_1^{(j)} \in \mathcal{E}\left(\widehat {m_1^{(j)}}, k_1 {S_1^{(j)}}^{-1} \right)}
f_{\mathrm{KL}}( e^{(j)}, m_0^{(j)}, m_1^{(j)}, S_0^{(j)}, S_1^{(j)}) \\
\mbox{subject to} & {e^{(j)}}^\top e^{(j)} = 1
\end{array} \label{optim-problem-oneDi3}
\end{equation}
where $\widehat {m_i^{(j)}} = { U^{(j)} }^\top  \widehat{m_i}$ and $S_i^{(j)} = { U^{(j)} }^\top S_i U^{(j)}$, $i=0,1$. That is,~\eqref{optim-problem-oneDi3} is simply an instance of~\eqref{op-KL-relaxation-p-eq-1} in the reduced dimensional space~${\mathbb R}^{n-j+1}$, for which we have developed a global solution.
Algorithm~\ref{greedy-algorithm} outlines the overall approach.
\vspace{-1.0em}
\begin{algorithm}
\caption{Greedy algorithm}
\label{alg:greedy}
\begin{algorithmic}[1]
\vspace{0.4mm}
\FOR {$j =1$ to $p$ }
\vspace{0.4mm}
    \STATE  Compute $U^{(j)} \in {\mathbb R}^{n \times ( n - j + 1 )}$ ($U^{(1)} := I_{n}$), an orthonormal basis for
    the orthogonal complement of the $j-1$ dimensional subspace  $\mathrm{span} \left\{ e_1, e_2, \ldots, e_{j-1} \right\}$
    \vspace{0.4mm}
    \STATE Compute the projected means and covariances $m_i^{(j)} = {U^{(j)}}^\top m_i$, $S_i^{(j)} = {U^{(j)}}^\top S_i U^{(j)}$ for $i = 0, 1$
    \vspace{0.4mm}
    \STATE
Compute $S^{(j)}=(S_0^{(j)})^{-1/2} S_{1}^{(j)} ({S_0}^{(j)})^{-1/2} $, $m^{(j)}=({S_0}^{(j)})^{-1/2}(m_1^{(j)}-m_0^{(j)})$, $M^{(j)}=m^{(j)} {m^{(j)}}^\top$
    \vspace{0.4mm}
    \STATE
        Solve \eqref{op-psi-KL} for $(M,S): = \left( M^{(j)}, S^{(j)} \right)$; find $e^{(j)} \in {\mathbb R}^{n-j+1}$ as: $e^{(j)}:=e^{\star}$, as in Lemma~\ref{lemma-first-step-KL}
    \vspace{0.4mm}
    \STATE
        Compute the $j$th column of~$E$ as $e_j = U^{(j)} e^{(j)}$
\vspace{0.4mm}
\ENDFOR
\end{algorithmic}
\label{greedy-algorithm}
\end{algorithm}
\subsubsection{Projection phase}
Relaxation phase~\ref{subsubs-relax-phase} produces a Stiefel matrix $E$.
Now, we project the matrix $E$ back to the set of $0/1$ selection matrices.
We remark that the objective function $f_{\mathrm{KL}}(\cdot)$ in eqn.~\eqref{KL-cost-fcn}
 depends on the matrix $E$
only through its range space; that is, $f_{\mathrm{KL}}\left(EQ\right)=f_{\mathrm{KL}}\left( E \right)$,
for any Stiefel matrix $E$ and for any orthogonal $p \times p$ matrix $Q$.
Thus, we choose the selection matrix $\widetilde E$ with the range space
closest to the range space of matrix $E$.
It can be shown (see~\cite{CDC-paper}) that $\widetilde E$ can be efficiently obtained as
follows: if $\left( j_1, j_2, \ldots, j_p\right)$ denote the indices of the largest
entries on the diagonal of $E E^\top$, then $\widetilde E =
\left[ h_{j_1} h_{j_2} \ldots h_{j_p} \right]$
where $h_j$ stands for the $j$-th column of the
identity matrix $I_n$. Thus, the Projection phase has very small computational cost.
\subsubsection{Refinement phase}
\label{sec-refinement}
Once the projection to the set of $0/1$ selection matrices is done and the matrix $\widetilde E$ is obtained,
we finalize our algorithm with a local maximization around~$\widetilde E$ to get ${E}^\star$ (see~\cite{Boyd, Fedorov}
 for very similar local searches.)
  Namely, for a given selection matrix $E$ in the
neighborhood of~$\widetilde E$, we find
\vspace{-0.7em}\[ f_{\mathrm{KL},\,\mathrm{worst}} (E):=  {\min}_{m_0 \in \mathcal{E} \left( \widehat m_0, k_0\, S_0^{-1} \right), m_1 \in \mathcal{E} \left( \widehat m_1, k_1\, S_1^{-1} \right)} f_\mathrm{KL}(E, m_0, m_1).\]
The procedure has $p$ steps. We start with the matrix $E:=\widetilde{E}$. In the first
 step, all columns of the current
selection matrix $E$ are fixed except the first one, which is viewed
as an optimization variable. The first column is swept through
all canonical vectors $h_j$, $j=1,\ldots,n$,
different from the remaining $p-1$ columns of $E$. After all possible
choices for the first column are tested, the column is
frozen to the choice that gives the maximal $f_{\mathrm{KL},\,\mathrm{worst}}$.
In the second step, this procedure is repeated for the second column, and so on,
 up to the $p$-th step; after the $p$-th step is done, we set $E^\star:=E$.

We obtain the quantity $f_{\mathrm{KL},\,\mathrm{worst}} (E)$ by first finding:
\begin{equation}
\label{op-KL-refinement}
\min_{
\substack{ m_0 \in \mathcal{E} \left( \widehat m_0, k_0\, S_0^{-1} \right), \\
m_1 \in \mathcal{E} \left( \widehat m_1, k_1\, S_1^{-1} \right)}}
(m_1-m_0)^\top E {(E^\top S_0 E)}^{-1} E^\top (m_1 - m_0),
\end{equation}
and then adding the remaining terms of the function $f_{\mathrm{KL}}(\cdot)$ that do not depend on $m_1$ and $m_0$ (see eqn. \eqref{KL-cost-fcn}).

Similarly as with the equivalence of~\eqref{optim-problem-oneDi2} and~\eqref{optim-problem-oneDi3}, it can be shown that the minimum in eqn. \eqref{op-KL-refinement} ($n$-dimensional problem) equals the following minimum (of corresponding $p$-dimensional problem):
\vspace{-0.7em}
\begin{equation}
\label{op-KL-refinement-lower-dim}
\min_{m_0' \in \mathcal{E} \left( Q^{\top} {(E^\top S_0 E)}^{-1/2} E^{\top} \widehat m_0, k_0\, I_p \right), m_1' \in \mathcal{E} \left( Q^{\top} {(E^\top S_0 E)}^{-1/2} E^{\top} \widehat m_1, k_1\, {\Lambda}^{-1} \right)}
\| m_1' - m_0' \|,
\end{equation}
where $\Lambda$ and $Q$ are, respectively, the matrix of eigenvalues  and the matrix of eigenvectors of \linebreak
$(E^\top S_0 E)^{-1/2} E^{\top}S_1E  (E^\top S_0 E)^{-1/2}$, and $\|\cdot\|$ denotes Euclidean norm. The problem~\eqref{op-KL-refinement-lower-dim} is a (convex) quadratically constrained quadratic problem (QCQP), and it can be solved with complexity $\O(p^3)$ (see, e.g., \cite{Nesterov-book}).
\vspace{-1.0em}
\subsection{Algorithm for the robust Chernoff-based selection: the R--C algorithm}
\label{subsec-R-C}
In this subsection, we present the algorithm R--C, the Chernoff based sensor selection under the presence of uncertainties. As mentioned previously, we adopt the same methodology for solving both~\eqref{op-C} and~\eqref{op-KL} and, consequently, the structure of R--C is the same as the one in R--KL. However, the problem~\eqref{op-C} is more difficult than~\eqref{op-KL}, due to the additional maximization over the parameter $s$. This will result in several specificities in R--C compared to R--KL. We present R--C by focusing on these specificities, phase by phase, whereas the overall structure remains the same as in R--KL.

The main difference between R--KL and R--C is in the Relaxation phase in the case $p=1$. As with the KL case, we solve the resulting Chernoff problem globally; we next explain a solution.
\subsubsection{Relaxation phase: case $p=1$: global solution}
Stiefel relaxation of~\eqref{op-C} for $p=1$ is given by:
\begin{equation}
\begin{array}[+]{ll}
\mbox{maximize} & \mbox{min}_{m_0 \in \mathcal{E} \left( \widehat m_0, k_0\, S_0^{-1} \right), m_1 \in \mathcal{E} \left( \widehat m_1, k_1\, S_1^{-1} \right)} \mbox{max}_{s\in [0,1]} f_\mathrm{C}(s, e, m_0, m_1) \\
\mbox{subject to} & e^\top e = 1,
\end{array}\label{op-C-relaxation-p-eq-1}
\end{equation}
where
\vspace{-1em}$$f_\mathrm{C}(s, e, m_0, m_1)=\frac{s(1-s)}{2}\frac{\left(e^\top (m_1 - m_0)\right)^2}{se^\top S_0 e +(1-s)e^\top S_1 e}- \frac{1}{2}\log \frac { (e^\top S_0 e)^s (e^\top S_1 e)^{1-s}}{se^\top S_0 e +(1-s)e^\top S_1 e}.$$
The first reformulation of~\eqref{op-C-relaxation-p-eq-1} that we make is the conversion of the minimax into maximin problem,
as Lemma~\ref{lemma-maxmin} explains.
\begin{lemma}
\label{lemma-maxmin}
Problem~\eqref{op-C-relaxation-p-eq-1} is equivalent to:
\begin{equation}
\begin{array}[+]{ll}
\mbox{maximize} & \mbox{max}_{e^\top e=1} \mbox{min}_{m_0 \in \mathcal{E} \left( \widehat m_0, k_0\, S_0^{-1} \right), m_1 \in \mathcal{E} \left( \widehat m_1, k_1\, S_1^{-1} \right)} f_\mathrm{C}(s, e, m_0, m_1) \\
\mbox{subject to} & s\in [0,1].
\end{array}\label{op-C-relaxation-p-eq-1-maxmin}
\end{equation}
\end{lemma}
\begin{proof}
Function $f_\mathrm{C}$ is convex w.r.t. $m_0$ and $m_1$, and concave w.r.t. $s$, and the constraint sets are compact and convex. Thus, the equivalence follows by Sion's minimax theorem (\cite{Simons}).
\end{proof}
Next, we focus on the inner maximization in~\eqref{op-C-relaxation-p-eq-1-maxmin}:
\begin{equation}
\begin{array}[+]{ll}
\mbox{maximize} & \mbox{min}_{m_0 \in \mathcal{E} \left( \widehat m_0, k_0\, S_0^{-1} \right), m_1 \in \mathcal{E} \left( \widehat m_1, k_1\, S_1^{-1} \right)} f_\mathrm{C}(s, e, m_0, m_1) \\
\mbox{subject to} & e^\top e=1
\end{array},\label{op-C-relaxation-p-eq-1-maxmin-inner-op}
\end{equation}
where $s \in [0,1]$ is fixed.
The following lemma is the counterpart of Lemma~\ref{lemma-first-step-KL}.
\begin{lemma}
\label{lemma-first-step-CS}
Suppose $(x^\star,\,y^\star)$ solves
\begin{equation}
\begin{array}[+]{ll}
\mbox{maximize} & \psi_{\mathrm{C}}(s,x,y) \\
\mbox{subject to} & (x,y) \in \mathcal{R}
\end{array} \label{op-psi-C}
\end{equation}
where
\vspace{-1em}
\begin{equation*}
\psi_{\mathrm{C}}(s,x,y) =\frac{s(1-s)}{2} \frac{\left\{ \left(\sqrt{y} - \frac{1}{\sqrt{k_1}}  \sqrt{ x } - \frac{1}{\sqrt{k_0}} \right)^{+} \right\}^2}
{s+(1-s)x} - \frac{1}{2} (1-s)\log x + \frac{1}{2} \log (s+(1-s)x),\\
 \end{equation*}
and $s \in [0,1]$.
Let $v^\star \in {\mathbb R}^n$ be an unit norm vector that generates $x^\star$ and $y^\star$, i.e., $x^\star={v^\star}^\top S v^\star$ and $y^\star={v^\star}^\top mm^\top v^\star$.
Then, $e^\star := S_0^{-1/2}v^\star/\|S_0^{-1/2}v^\star\|$ solves~\eqref{op-C-relaxation-p-eq-1-maxmin-inner-op}.
\end{lemma}
\begin{proof} The proof is similar to the proof of Lemma~\ref{lemma-first-step-KL} and is omitted.
\end{proof}
By similar analysis as in subsection~\ref{subsubs-relax-phase}, it can be shown that~\eqref{op-psi-C} can be solved by parametrization of the boundary of $\mathcal {R}$, given in eqn.~\eqref{def-R}. Thus, for fixed $s$, the algorithm that solves~\eqref{op-C-relaxation-p-eq-1-maxmin-inner-op} is the same as the one that solves~\eqref{op-KL-relaxation-p-eq-1},
 except that $\psi_{\mathrm{KL}}(x,y)$ is replaced by $\psi_{\mathrm{C}}(s,x,y)$. Then, the function $\psi_{\mathrm{C}}(s, x^\star(s), y^\star(s))$ can be evaluated using this algorithm and problem~\eqref{op-C-relaxation-p-eq-1}
 is solvable by, e.g., a grid search on the interval $[0,1]$.

The projection phase of R--C is the same as the projection phase of R--KL and the steps in the
refinement phase of R--C are the same as the ones in the refinement phase of R--KL (with $f_{\mathrm{KL}}(E,m_0,m_1)$ replaced by $\max_{s \in [0,1]} f_{\mathrm{C}}(s,E,m_0,m_1)$). Similarly as in~\ref{sec-refinement}, in the refinement phase, for a given selection matrix $E$, we have to find
\[ f_{\mathrm{C},\,\mathrm{worst}} (E):=  {\min}_{m_0 \in \mathcal{E} \left(\widehat m_0, k_0\, S_0^{-1} \right), m_1 \in \mathcal{E} \left( \widehat m_1, k_1\, S_1^{-1} \right)} \max_{s\in [0,1]} f_\mathrm{C}(E, m_0, m_1).\] Applying again the minimax theorem, we first exchange the order of min and max in $f_{\mathrm{C},\,\mathrm{worst}}$; then, for fixed $s$, we find
\begin{equation}
\label{op-C-refinement}
\min_
{\substack{ m_0 \in \mathcal{E} \left( \widehat m_0, k_0\, S_0^{-1} \right),  \\
m_1 \in \mathcal{E} \left( \widehat m_1, k_1\, S_1^{-1} \right)}}
(m_1-m_0)^\top E\left(sE^\top S_0 E + (1-s)E^\top S_1 E\right)^{-1} E^\top (m_1 - m_0),
\end{equation}
by solving the equivalent $p$-dimensional QCQP~\eqref{op-C-refinement-lower-dim} (with equal minimum):
\vspace{-10pt}
\begin{equation}
\label{op-C-refinement-lower-dim}
\min_{
\substack{m_0' \in \mathcal{E} \left( Q^\top (E^\top S_0 E)^{-1/2} E^{\top} \widehat m_0, k_0\, I_p \right),\\
m_1' \in \mathcal{E} \left( Q^\top (E^\top S_0 E)^{-1/2} E^{\top} \widehat m_1, k_1\, {\Lambda}^{-1}\right) }}
(m_1' - m_0')^\top{(sI_p +(1-s)\Lambda )}^{-1} (m_1' - m_0').
\end{equation}
Finally, we compute $f_{\mathrm{C},\,\mathrm{worst}} (E)$ by the bisection method on parameter $s$.
\vspace{-1em}
\subsection{Complexities of R--KL and R--C}
\label{subsec-complexities-RKL-RC}
The complexity of both R--KL and R--C algorithm is $\O(n^3p+np^4)$, although the hidden constant in R--C is larger
 than the one in R--KL. The least computational effort is required for the projection phase $2$, which for both R--C and R--KL is $\O(n^2)$ and is dominated by complexities of the other two phases. It can be shown that Phase 1 has complexity $\O(n^3p)$ and Phase 3 has complexity $\O(np^4)$.
\vspace{-0.5em}
\section{Sensor selection algorithms: No uncertainties case}
\label{section-no-uncert}
In this section, we address a special, yet important case, when there are no uncertainties in the mean vectors and the problems~\eqref{op-KL} and~\eqref{op-C} simplify by dropping the inner minimizations.
We first remark that algorithms R--KL and R--C can readily solve the simplified versions of~\eqref{op-KL} and~\eqref{op-C}. However, we derive in this section a more efficient algorithm. We exploit the structure of the problem and the knowledge
of exact distribution parameters $m_i$, $i=0,1$ (more specifically, their difference $m_1-m_0$) to
 reduce the computational load of the relaxation phase of R--KL
  and R--C algorithm, while keeping the second and the third phase the same\footnote{We remark that the refinement phase of R--C and R--KL simplifies significantly in the no uncertainties case, as in this case computing $f_{\mathrm{KL},\,\mathrm{worst}}(E)$ boils down to computing $f_{\mathrm{KL}}(E)$, i.e. there is no need to solve intermediate minimization problems (see ~\eqref{op-KL-refinement} and~\eqref{op-C-refinement}).}. The key to reducing the complexity of the relaxation phase is a
  \emph{simple, analytic solution} of the relaxed, Stiefel problem, in the special case of equal mean values.
 We refer to the overall simplified algorithms as MD--KL (mean-difference based KL algorithm), and MD--C (mean-difference based C algorithm).
\vspace{-1em}
\subsection{Kullback-Leibler based selection without uncertainties: The mean-difference KL algorithm (MD--KL)}
\label{subsect-MD-KL}
In this subsection, we only explain the relaxation phase of MD--KL, as the other two phases are the same as in R--KL.
We first consider a special case of equal mean values of the problem~\eqref{eq-KL-op-relaxation} (without inner minimization), and we show that this problem has a simple analytic solution. Based on the solution for the equal means, we derive an algorithm that solves the general case.

\subsubsection{Relaxation phase: the case $m_0=m_1$}
\label{subsubsec-KL-eq-means}
Consider the problem~\eqref{eq-KL-op-relaxation} when there is no uncertainty in the mean values (uncertainty ellipsoids shrink to a point, by letting $k_0, \, k_1=\infty$, and inner minimization drops from the problem).
We first remark that we can write~\eqref{eq-KL-op-relaxation} when $k_i=\infty$, $m_1=m_0$, in the following form:
\begin{equation}
\begin{array}[+]{ll}
\mbox{maximize} & \frac{1}{2}\left(\mathrm{tr} (P^\top SP)-\log \left|P^\top S P\right|-p\right) \\
\mbox{subject to} & P^\top P = I_{p},
\end{array}\label{op-KL-no-uncert-relaxation-reformulated}
\end{equation}
where $S=S_0^{-1/2} S_1 S_0^{-1/2}$. This equivalence can be shown by noting that the constraint $E^\top E = I_{p}$
in~\eqref{eq-KL-op-relaxation} can be replaced by $E^\top S_0 E = I_{p}$, and by introducing the new variable $P=S_0^{1/2}E$.

The objective function in~\eqref{op-KL-no-uncert-relaxation-reformulated} can be further simplified to
 $\sum_{i=1}^p   \phi_{\mathrm{KL}}\left(\lambda_i\left(P^\top SP \right)\right)$, where $\phi_{\mathrm{KL}}(x)=x-\log x-1$ and $\lambda_i$ denotes the $i$-th largest eigenvalue. Invoking Poincar\'{e} separation theorem (see~\cite{MatrixAnalysis}), we show in the Appendix~\ref{App-Equal-Cov-KL} that the solution $P^\star$ of~\eqref{op-KL-no-uncert-relaxation-reformulated}
is given by a set of $p$ orthonormal eigenvectors of $S$ corresponding to the $p$ eigenvalues that give the highest objective $\phi_{\mathrm{KL}}(\cdot)$. It turns out
 that we do not need to check all $n \choose p$ eigenvector subsets, but only at most $p+1$ of them.
 We now give the procedure in Algorithm~\ref{alg:KL-eq-means} to choose the optimal subset of eigenvectors (see Appendix~\ref{App-Equal-Cov-KL} for the proof).
\begin{algorithm}
\caption{Procedure for solving~\eqref{op-KL-no-uncert-relaxation-reformulated} when $m_1=m_0$}
\label{alg:KL-eq-means}
\begin{algorithmic}[1]
\vspace{0.4mm}
\STATE Set $\phi_{\mathrm{KL}}^{\star}=0$
\FOR {$j =0$ to $p$ }
\vspace{0.4mm}
    \STATE  $x=(\lambda_1(S),\, \ldots,\, \lambda_{j}(S),\; \lambda_{n-p+j+1}(S),\,\ldots,\, \lambda_n(S))$
    \vspace{0.4mm}
    \STATE Compute $\phi =\sum_{i=1}^p \phi_{\mathrm{KL}}(x_i)$
    \vspace{0.4mm}
    \STATE
        if $\phi>\phi_{\mathrm{KL}}^{\star}$, then $j^{\star}=j$, $\phi_{\mathrm{KL}}^{\star}=\phi$
\vspace{0.4mm}
\ENDFOR
\STATE $x^{\star}=(\lambda_1(S),\, \ldots,\, \lambda_{j^{\star}}(S),\; \lambda_{n-p+j^{\star}+1}(S),\,\ldots,\, \lambda_n(S))$; \\
$P^{\star}$ is the set of eigenvectors of $S$ corresponding to the eigenvalues of $S$ given in $x^{\star}$.
\end{algorithmic}
\end{algorithm}

We give the intuition behind the solution of~\eqref{op-KL-no-uncert-relaxation-reformulated}. Recall that the matrix $E$ is chosen such that the projection of the covariance matrix $S_0$ equals $E^\top S_0 E=I_p$. Then, the projection of $S_1$, $E^\top S_1 E$, equals $P^\top S P$. Thus, the further from point $1$ are the eigenvalues of $P^\top S P$, the better are the projected distributions separated. The function $\phi_{\mathrm{KL}}(\cdot)$ measures the distance from $1$ in this sense.

\subsubsection{Relaxation phase: general case}
\label{subsubsec-KL-general-case}
 The main idea behind the relaxation phase of MD--KL is as follows: set one column of the solution Stiefel matrix $E$ in the direction of the vector $m_1-m_0$, i.e., in the direction
   of the difference of the distribution means. The remaining $p-1$ columns of $E$ are then
    obtained in the following way: we project the distribution parameters $m_i,\,S_i$, $i=0,1,$ to the orthogonal complement of
    $m_1-m_0$, and then solve an ($p-1$ dimensional) instance of~\eqref{eq-KL-op-relaxation} when $k_i=\infty$,
    with the projected distribution parameters. This $p-1$-dimensional instance of~\eqref{eq-KL-op-relaxation} is
     in fact the special one, with equal means, and is
     hence very efficiently solved by procedure given in Algorithm~\eqref{alg:KL-eq-means}. The relaxation phase
      of MD--KL is summarized in Algorithm~\ref{alg:MD--KL-alg}. We give the intuition behind
       the choice of $m_1-m_0$ as the direction of the first column of $E$:
       the Euclidian distance between the means $E^\top m_1$ and $E^\top m_0$
        of the projected distributions is maximal possible (and equal to $\|m_1-m_0\|$),
         when one column of $E$ lies in the direction of $m_1-m_0$.
\begin{algorithm}
\caption{MD--KL algorithm: Relaxation phase}
\label{alg:MD--KL-alg}
\begin{algorithmic}[1]
\vspace{0.4mm}
    \STATE  Set $e_1:=\left( m_1-m_0\right) / \|m_1-m_0\|$
    \STATE  Compute $U \in {\mathbb R}^{n \times ( n -1 )}$, an orthonormal basis for the orthogonal complement of $e_1$
    \STATE  Compute $S^\prime = \left( U^\top\, S_0 \, U \right)^{-1/2} \,U^\top\, S_1 \, U \, \left( U^\top\, S_0 \, U \right)^{-1/2} $
    \STATE  Find $P$ as in Algorithm~\ref{alg:KL-eq-means} for $S:=S^\prime$, and $p:=p-1$
    \STATE  $[e_2,\,e_3,...,e_p]:=U\,S_0^{-1/2}\,P$, $E:=[e_1,\,e_2,...,e_p]$
\end{algorithmic}
\end{algorithm}

\subsection{Chernoff based selection without uncertainties: The mean-difference C algorithm (MD--C)}
\label{subsect-MD-C}

\subsubsection{Relaxation phase: the case $m_0=m_1$}
\label{subsubsec-C-eq-means}
The counterpart of problem~\eqref{op-KL-no-uncert-relaxation-reformulated} for the C criterion is the following:
  \begin{equation}
\begin{array}[+]{ll}
\mbox{maximize} & \mbox{max}_{s\in [0,1]} \, \,
-\log \frac{\left|P^{\top} SP \right|^{1-s}} {\left|sI +(1-s)P^{\top} SP\right|}\\
\mbox{subject to} & P^{\top} P = I_{p}
\end{array},\label{opt-C-reformulated}
\end{equation}
The objective in~\eqref{opt-C-reformulated} can be
written as $\sum_{i=1}^{p}\phi_{\mathrm{C}} \left(s,\lambda_i\left(P^{\top} SP\right)\right)$, where $\phi_{\mathrm{C}}(s,x):=\log (s+(1-s)x)-(1-s)\log x$. Now, the problem~\eqref{opt-C-reformulated} ($m_1=m_0$) can be solved
by a method similar to the one in~\ref{subsubsec-KL-eq-means}. Namely, 1) by invoking Poincar\'{e} separation theorem; 2) by using concavity of $\phi_{\mathrm{C}}(\cdot,x)$; and 3) by using unimodularity
  of $\phi_{\mathrm{C}}(s,\cdot)$, it can be shown that
    the solution to~\eqref{opt-C-reformulated} ($m_1=m_0$) can be obtained by the
    procedure given in Algorithm~\ref{alg:equal-means-Chernoff}.
\begin{algorithm}
\caption{Procedure for solving~\eqref{opt-C-reformulated} when $m_1=m_0$}
\label{alg:equal-means-Chernoff}
\begin{algorithmic}[1]
\vspace{0.4mm}
\STATE Set $\phi_{\mathrm{C}}^{\star}=0$
\FOR {$j =0$ to $p$ }
\vspace{0.4mm}
    \STATE  $x=(\lambda_1(S),\, \ldots ,\, \lambda_{j}(S),\; \lambda_{n-p+j+1}(S),\,\ldots,\, \lambda_n(S))$
    \vspace{0.4mm}
    \STATE Find $s(x) \in \mathrm{arg}\,\max_{s \in [0,1]} \sum_{i=1}^p \phi_{\mathrm{C}}(s,x_i)$ by Newton method
    \STATE Compute $\phi =\sum_{i=1}^p \phi_{\mathrm{C}}(s(x),x_i)$
    \vspace{0.4mm}
    \STATE
        if $\phi>\phi_{\mathrm{C}}^{\star}$, then $j^{\star}=j$, $s^{\star}=s(x)$, $\phi_{\mathrm{C}}^{\star}=\phi$
\vspace{0.4mm}
\ENDFOR
\STATE $x^{\star}=(\lambda_1(S),\, \ldots,\, \lambda_{j^{\star}}(S),\; \lambda_{n-p+j^{\star}+1}(S),\,\ldots,\, \lambda_n(S))$;
 $s^{\star}=s^{\star}$;\\
 $P^{\star}$ is the set of orthonormal eigenvectors of $S$ corresponding to the eigenvalues of $S$ given in $x^{\star}$.
\end{algorithmic}
\end{algorithm}

General case of the relaxation phase of MD--C algorithm is the same as the one in MD--KL given in~\ref{alg:MD--KL-alg}, except that, in step 4), it calls procedure given by Algorithm~\ref{alg:equal-means-Chernoff}.
\vspace{-1.0em}
\subsection{Complexities of MD--KL and MD--C}
\label{subsec-complexities-MDKL-MDC}
We briefly comment on the complexity of MD--KL and MD--C. The complexity of both MD--KL and MD--C is $\O(n^3+np^3)$, although the hidden constant in MD--C is larger than the one in MD--C. The main computational burden in the relaxation phase
of MD--KL is computation of the orthogonal complement $U$ of $m_1-m_0$, and subsequent eigenvalue decomposition of the supplementary matrix $S$ (see Algorithm~3), which is in total of order $\O(n^3)$. The relaxation phase of MD--C requires more computational effort; besides finding $U$ and the eigenvalue decomposition of $S$, each of the $p-1$ steps in Algorithm $4$ requires finding optimal $s$ by Newton method (contrary to just evaluating the KL distance for a given choice of $p-1$ eigenvalues of $S$ with Algorithm $2$). However, the number of operations in $p-1$ Newton runs is still dominated by the number of operations to find $U$. Therefore, the complexity of the relaxation phase of MD--C is $\O(n^3)$. Finally, it can be shown that, for both MD--KL and MD--C, the refinement phase is of complexity $\O(np^3)$.

\section{Numerical studies: Testing the optimization criteria}
\label{subsect-Test-criteria}
This subsection tests how good are the Kullback-Leibler and the Chernoff distance as
optimality criteria for sensor selection. To this end,
we want to compare the sensor selections that optimize the KL and C distances with: 1) the sensor selection
 that minimizes the Bayes probability of error (Bayes optimality); 2)
 the sensor selection that minimizes the probability
 of miss subject to a given
 probability of false alarm (Neyman-Pearson optimality).
 We find numerically the Bayes optimal and the Neyman-Pearson
 optimal sensor selections by Monte Carlo simulations.
 With respect to Bayes optimality,
  for each possible (out of $n \choose p$) sensor selection,
   we estimate the Bayes
   probability of error $P_{\mathrm{e}}$ by
    Monte Carlo simulations with $100,000$
    instantiations of the maximum-likelihood detector tests (with zero treshold), with equal prior probabilities. With respect to
    Neyman-Pearson optimality,
    for the fixed probability of false alarm $P_{\mathrm{FA}}$,
    we find the sensor selection
     that maximizes the probability of detection $P_{\mathrm{D}}$.
     This is achieved by estimating the receiver operating characteristic (ROC)
      curve (by Monte Carlo simulations with $20,000$ instantiations of the likelihood ratio tests) in the neighborhood of $P_{\mathrm{FA}}$ in a fine grid, and then
       interpolating the ROC curve (i.e., $P_{\mathrm{D}}$) at the desired point $P_{\mathrm{FA}}$.
        This is done for each possible selection and
        the selection with maximal obtained $P_{\mathrm{D}}$ is set as Neyman-Pearson optimal.

Table~I shows the Bayes probability of error for: 1) the sensor selection that
maximizes KL distance (KL); 2) the sensor selection that
maximizes C distance (C); 3) the best sensor selection (that minimizes $P_{\mathrm{e}}$);
 4) the worst sensor selection (that maximizes $P_{\mathrm{e}}$);
  and 5) the average $P_{\mathrm{e}}$ over all selections.
   Table~I (left) is for $n=12$ and $p=2,3,4,5$ and Table~I (right)
    is for $n=15$ and $p=2,3,4,6$. We can see that C selection matches the best selection.
     KL selection is in many cases very close or equal to C and best selections in $P_{\mathrm{e}}$.
      As could be predicted by theory (Chernoff lemma), the C selection is better than the KL selection in terms of $P_{\mathrm{e}}$.

Comparison of: 1) C, KL and best selections; with 2) worst and average selections
justifies the sensor selection problem; namely, by finding the optimal selection,
$P_{\mathrm{e}}$ can be for an order of magnitude smaller than for the average selection. (See Table~1 and Figure~1-right.)

Table~{II} shows the probability of detection for: 1) the KL-optimal selection (KL);
  2) the C-optimal selection (C); and 3) the Neyman-Pearson (NP) optimal selection, for $n=12$, $p=2,3,4,5$. We can see that, for $p=2$ and $p=3$, both KL and C selections match the NP optimal selections, and for $p=3$ and $p=4$, $P_{\mathrm{D}}$ for KL and C selections is at most $3.5\%$ from the optimum ($p=4$, $P_{\mathrm{D}}=0.05$).

\begin{figure}[thpb]
      \centering
      \includegraphics[height=2.16in,width=3.06in]{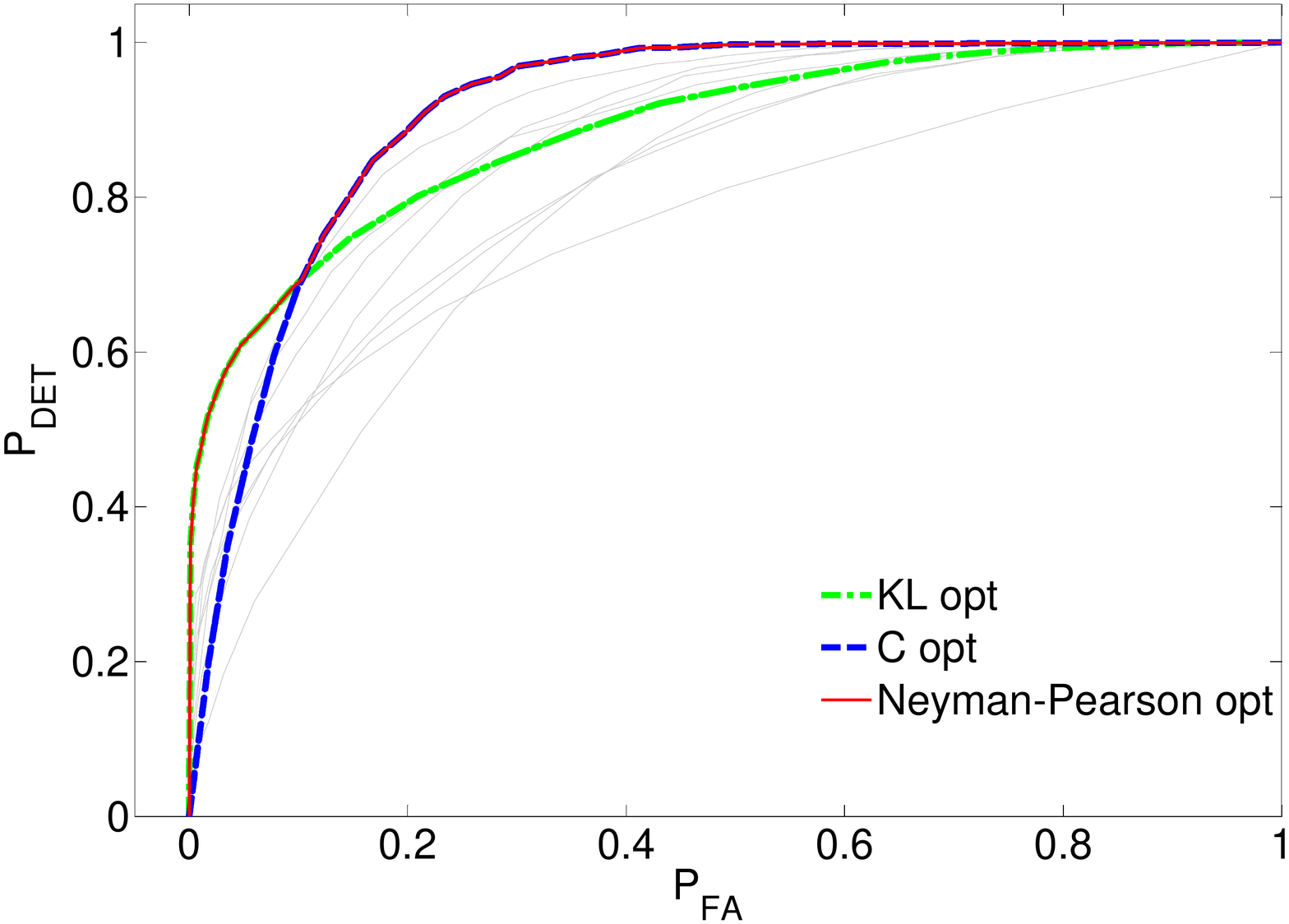}
      \includegraphics[height=2.16in,width=3.06in]{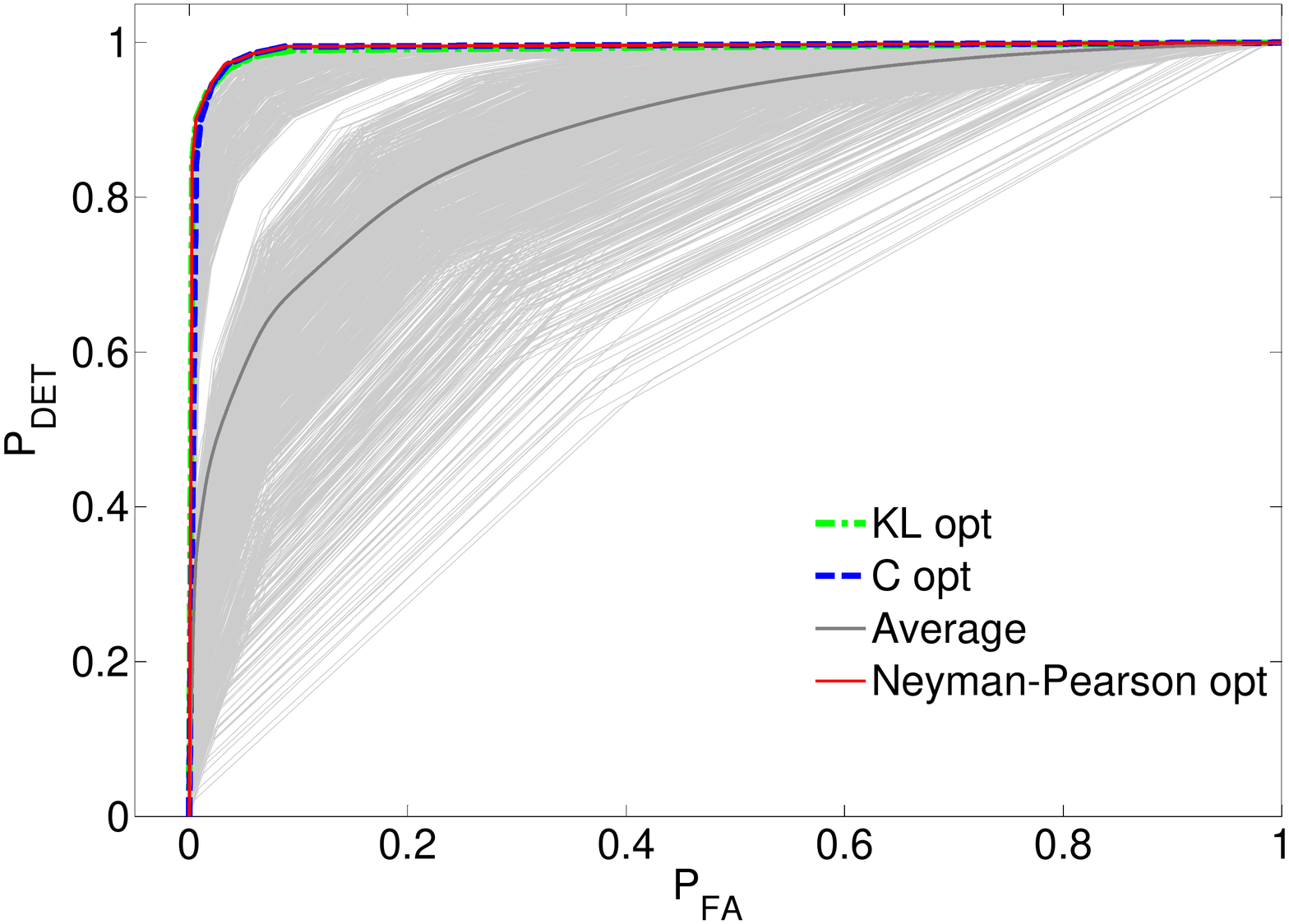}
      \caption{ Left: $n=5$, $p=2$; ROC curves for all possible sensor selections (gray); the ROC curves for the KL-optimal and the C-optimal sensor selections, as well as the Neyman-Pearson optimal ``envelope" are represented in different color. Right:
       $n=15$, $p=4$, Figure plots the same ROC curves as on the left; in addition, the Figure plots the average ROC among all sensor selections.}\label{fig-ROC}
\vspace{-15pt}
\end{figure}

    Figure~\ref{fig-ROC} (left) plots the ROC curves
    for all possible selections, for $n=5$ and $p=2$. We plot
     the ROC curves for: 1) KL-optimal selection; 2) C-optimal selection; 3) the pointwise
     envelope of all possible curves (Neyman-Pearson optimal). Remark that the Neyman-Pearson optimal
     curve (envelope) is not obtained for a single selection; in different regions,
     it corresponds to different selections. We can see that for lower values
     of $P_{\mathrm{FA}}$, KL selection is optimal; for higher values of $P_{\mathrm{D}}$,
     C-selection is optimal.

     Figure~\ref{fig-ROC} plots the ROC curve for
     all possible ${ n \choose p }= 1365$ selections for a larger example, with $n=15$ and $p=4$.
     Interestingly, we can see that the KL and C selections are
     very close to the optimum, in whole range of $P_{\mathrm{FA}}$. In addition, we
      plot the average of the ROC curves (pointwise average of $P_{\mathrm{D}}$ for each fixed $P_{\mathrm{FA}}$).
       This average curve thus represents what performance would be, on average, achieved, if
        we choose a subset of sensors uniformly at random. We can see that there is a a large gain
         of the C and KL selections over this average curve; thus, selecting the optimal, rather than random subset of sensors, provides large performance gain.

         Finally, we remark that, in extensive simulations, we observe
          similar behavior as in representative Tables~{I}~and~{II}, and Figure~\ref{fig-ROC} (left and right). That is, the KL and C selections are very close to optimal and even equal to optimal in certain range of $P_{\mathrm{FA}}$.
          We also report that C-selection is generally better than KL for large $P_{\mathrm{D}}$'s (upper right
           part of the ROC curve,) while KL is generally better for low $P_{\mathrm{FA}}$ (lower left in the ROC). Improvement
            of KL over C for low $P_{\mathrm{FA}}$ is smaller than the improvement of C over KL for large $P_{\mathrm{D}}$.
\newcommand{\ra}[1]{\renewcommand{\arraystretch}{#1}}
 \setlength{\abovecaptionskip}{-0ex}
\begin{table}
\centering
\caption{$P_{\mathrm{e}}$ for KL, C and optimal selection; Left: $n=12$, $p=2,3,4,5$;
 Right: $n=15$, $p=2,3,4,6$}
\ra{1.2}
\begin{minipage}{3.2in}
\begin{tabular}{@{}rrrrr@{}}\toprule
$P_{\mathrm{e}}$ & $p=2$ & $p=3$ & $p=4$ & $p=5$\\
\midrule
KL &  0.100 & 0.078 & 0.052 & 0.046\\
C & 0.100 & 0.078 & 0.052 & 0.043\\
Bayes-best & 0.100 & 0.078 & 0.052 & 0.043\\
worst & 0.457 & 0.439 & 0.396 & 0.319\\
average & 0.275 & 0.216 & 0.170 & 0.134\\
\bottomrule
\end{tabular}
\end{minipage}
\begin{minipage}{3.2in}
\ra{1.2}
\begin{tabular}{@{}rrrrr@{}}\toprule
$P_{\mathrm{e}}$ & $p=2$ & $p=3$ & $p=4$ & $p=6$\\
\midrule
KL &  0.086 & 0.061 & 0.037 & 0.022\\
C & 0.085 & 0.051 & 0.029 & 0.012\\
Bayes-best & 0.085 & 0.051 & 0.029 & 0.012\\
worst & 0.480 & 0.440 & 0.396 & 0.311\\
average & 0.240 & 0.180 & 0.136 & 0.077\\
\bottomrule
\end{tabular}
\end{minipage}
\end{table}
 \setlength{\belowcaptionskip}{-1ex}

 \setlength{\abovecaptionskip}{-1ex}
\begin{table}\centering
\caption{$P_{\mathrm{D}}$ for $P_{\mathrm{FA}}=[0.005\,0.03\, 0.1]$, $n=15$, $p=2,3,4,5$ for KL, C and optimal selection}
\ra{1.2}
\begin{tabular}{@{}rlllllllllll@{}}\toprule
 & \multicolumn{3}{c}{$P_{\mathrm{FA}}=0.005$} & \phantom{abc}& \multicolumn{3}{c}{$P_{\mathrm{FA}}=0.03$} &
\phantom{abc} & \multicolumn{3}{c}{$P_{\mathrm{FA}}=0.1$}\\
\cmidrule{2-4} \cmidrule{6-8} \cmidrule{10-12}
$P_{\mathrm{D}}\;\;\;\,$ & KL & C & NP && KL & C & NP && KL & C & NP\\
\hline
$p=2\;\;\;\;$ & 0.720 & 0.720 & 0.720 &&  0.889 & 0.889 & 0.889 && 0.961 &  0.961 & 0.961\\
$p=3\;\;\;\;$ & 0.812 & 0.812 & 0.812 &&  0.941 & 0.941 & 0.941 && 0.985 &  0.985 & 0.985\\
$p=4\;\;\;\;$ & 0.868 & 0.838 & 0.868 &&  0.962 & 0.967 & 0.970 && 0.990 &  0.995 & 0.995\\
$p=5\;\;\;\;$ & 0.891 & 0.898 & 0.903 &&  0.968 & 0.981 & 0.983 && 0.991 &  0.994 & 0.996\\
\bottomrule
\end{tabular}
\end{table}
 \setlength{\belowcaptionskip}{-1ex}
\section{Numerical studies: Testing the algorithms}
\label{subsect-test-alg}
This subsection tests our algorithms for solving
 the sensor selection problem. Subsection~\ref{subsect-test-alg-uncert}
 shows that algorithms R--KL and R--C show near optimal performance in the presence of uncertainties (i.e., for solving the
  problems~\eqref{op-KL}~and~\eqref{op-C}), Subsection (\ref{subsect-test-alg-no-uncert})
   shows that R--KL and R--C have near optimal performance when also applied to the case of no uncertainties (i.e., when $k_0=k_1=\infty$ in~\eqref{op-KL}~and~\eqref{op-C}). This subsection also shows
   that MD--KL (resp. MD--C) has comparable performance to R--KL (resp. R--C), while reducing the computational time.

For smaller problem instances, i.e., for smaller values of $n$ and $p$, we compare
   the solutions produced by our algorithms with the optimum obtained by exhaustive search.
   For larger examples, when the optimum is infeasible to
   compute, we generate randomly a number
   of sensor selections; then, we compare the selection obtained by our algorithms
    with the best among generated random selections.

For a fixed size of the problem (for fixed pair $(n,p)$),
we generate randomly $K_{\mathrm{exper}}$ instances of the problem parameters ($m_i,S_i$, $i=0,1$); for the
case with uncertainties, $K_{\mathrm{exper}}=50$, and for the case
 without uncertainties, $K_{\mathrm{exper}}=200$. In the case with uncertainties, parameters $k_0$ and $k_1$ are chosen such that the norm of the mean vector drift does not exceed $0.15\|m_1-m_0\|$, i.e., we set them as $k_i=\left|S_i\right|/\left(0.15 \|m_1-m_0\|\right)^2,\,\,i=0,1$. Based on $K_{\mathrm{exper}}$
 solved problems of fixed size $(n,p)$, we create statistics on how our algorithms behave for given size of the problem.

\subsection{Testing the R--KL and R--C algorithms: The case with uncertainties}
\label{subsect-test-alg-uncert}

\mypar{Testing the robustness against the uncertainty in distribution means}
Table~{III} (left) shows performance of the R--KL and R--C algorithms, for solving~\eqref{op-KL} and~\eqref{op-C}, respectively. We evaluate the optimal solutions $f^{\mathrm{OPT-KL}}$ and $f^{\mathrm{OPT-C}}$ by exhaustive search. We then compute the ratio $r_{\mathrm{R-KL}}:=\frac{f^{\mathrm{R-KL}}}{f^{\mathrm{OPT-KL}}}$ (resp. $r_{\mathrm{R-C}}:=\frac{f^{\mathrm{R-C}}}{f^{\mathrm{OPT-C}}}$) that says how close is the solution value
 obtained by R--KL (resp. R--C) to optimum. Table~{III} (left) shows the results for
 $n=10, 12, 15$ and $p=3$. We present the maximum (max), the average (avg), the minimum (min),
 and the standard deviation (dev) of the ratio $r_{\mathrm{R-KL}}$ over $K_{\mathrm{exper}}=50$
  problem instances, for each pair $(n,p)$.
We can see that both R--KL and R--C
show very good performance; the R--C shows better performance than
R--KL. The maximal value of both $r_{\mathrm{R-KL}}$ and $r_{\mathrm{R-C}}$ in all the examples is equal to $1$. With the R--KL algorithm,
the average value of $r_{\mathrm{R-KL}}$ is always above $91.8\%$, and the minimum value
is always above $51\%$;
with the R--C algorithm, the average of $r_{\mathrm{R-C}}$ is always above $96\%$, and
the minimum is always above $59\%$.

Table~{III} (right) shows performance of R--KL and R--C for a larger example $n=50$, $p=5$, for which it was infeasible to find the optimal solution by exhaustive search. We thus randomly generate $2500$
sensor selections, and we evaluate the quantity $f^{\mathrm{BEST-RAND-KL}}$--the
maximal objective function in~\eqref{op-KL} over all
 $2500$ randomly generated selections.
 Define the ratios $\rho_{\mathrm{R-KL}}:=\frac{f^{\mathrm{R-KL}}}{f^{\mathrm{BEST-RAND-KL}}}$. (Analogously define the corresponding
   ratios for the Chernoff-based sensor selection.) We report that, to obtain $f^{\mathrm{BEST-RAND-KL}}$ (for $2500$ $50\times 5$ selection matrices) it takes about $10$ times longer than for R--KL algorithm to find a solution. Also, to obtain $f^{\mathrm{BEST-RAND-C}}$ it takes about $6$ times longer than for R--C algorithm to find a solution. From Table~{III} (right) we can see that R--KL in many cases outperforms random strategy: with significant savings in time ($90\%$) it gives a $26\%$ larger objective function on average. R--C always stays above the random strategy in terms of the objective function, with computational savings of $83\%$.
 \setlength{\abovecaptionskip}{-1ex}
\begin{table}
\centering
\caption{ Left: Statistics for $r_{\mathrm{R-KL}}$ and $r_{\mathrm{R-C}}$, smaller examples: $n=10, 12,15$, $p=3$, $K_{\mathrm{exper}}=50$;
 Right: Statistics for $\rho_{\mathrm{R-KL}}$ and $\rho_{\mathrm{R-C}}$, $n=50$, $p=5$, $K_{\mathrm{exper}}=50$}
\begin{minipage}{3.2in}
\ra{1.2}
\begin{tabular}{@{}rcccc@{}}\toprule
& $\mathrm{max}$ & $\mathrm{avg}$ & $\min$ & $\mathrm{dev}$\\
\midrule
\vspace*{-.4cm}\\
$r_{\mathrm{R-KL}}$\\
$n=10$ & 1.000 & 0.964 & 0.606  & 0.090
\\
$n=12$ & 1.000 & 0.918  & 0.551 & 0.140
\\
$n=15$ & 1.000 & 0.939 & 0.512 &  0.118
\\
\vspace*{-.4cm}\\
$r_{\mathrm{R-C}}$\\
$n=10$ & 1.000 &  0.981 & 0.7862 & 0.051
\\
$n=12$ & 1.000 &  0.982 & 0.834  & 0.037
\\
$n=15$ & 1.000 & 0.961 & 0.595 & 0.095
\\
\bottomrule
\end{tabular}
\end{minipage}
\begin{minipage}{3.2in}
\ra{1.2}
\begin{tabular}{@{}rcccc@{}}\toprule
& $\mathrm{max}$ & $\mathrm{avg}$ & $\min$ & $\mathrm{dev}$\\
\midrule
\vspace*{-.25cm}\\
$\rho_{\mathrm{R-KL}}$ & 1.777 & 1.267 & 0.817  & 0.191\\
$\rho_{\mathrm{R-C}}$ & 1.868 &  1.277 & 1.005 & 0.166\\
\bottomrule
\end{tabular}
\end{minipage}
\end{table}
\setlength{\belowcaptionskip}{-1ex}

\subsection{Testing the algorithms: The case without uncertainties}
\label{subsect-test-alg-no-uncert}

$\,$

\mypar{Smaller examples: Comparison with the global optimum} Table~{IV} shows performance of the R--KL and MD--KL algorithms for solving~\eqref{op-KL} in the case without uncertainties ($k_0=k_1=\infty$); table~{V} shows performance of the R--C and MD--C algorithms for solving the corresponding Chernoff problem~\eqref{op-KL}. For fixed $n$ and $p$, we generate $K_{\mathrm{exper}}=200$ sets of problem parameters $m_i, S_i, i=0,1$; for
 each fixed $m_i, S_i, i=0,1$, we evaluate the ratios $r_{\mathrm{R-KL}}$, $r_{\mathrm{MD-KL}}$, $r_{\mathrm{R-C}}$, and $r_{\mathrm{MD-C}}$. Table~{IV} shows the maximum (max), the average (avg) and the minimum (min) of the quantities
   $r_{\mathrm{R-KL}}$ and $r_{\mathrm{MD-KL}}$
    over all $K_{\mathrm{exper}}=200$ experiments. The standard deviation in these experiments varies from $0.03$ to $0.06$.
   We can see from the table that
     the maximum of both $r_{\mathrm{R-KL}}$ and $r_{\mathrm{MD-KL}}$
      is always $1$. Table~{IV} demonstrates very good performance of
       both R--KL and MD--KL algorithms. On average, $r_{\mathrm{R-KL}}$ and $r_{\mathrm{MD-KL}}$
       are always (for each pair $n, p$)
        above $97.7\%$; the minimum is always above $67.2\%$.
        We see that R--KL and MD--KL algorithms
        have comparable performance with respect to (near)optimality, while MD--KL has smaller
         computational cost (We note, however, that MD--KL does not apply for the
         case with uncertainties in distribution means.)

Table~{V} shows very good performance of the R--C algorithm and
the MD--C algorithm. As with the case of KL algorithms, the maximum is always $1$. The average of both quantities
$r_{\mathrm{R-C}}$ and $r_{\mathrm{MD-C}}$
 is always above $99.2\%$, and the minimum is always
 above $78.9\%$. Standard deviation is smaller than the one in R--KL and varies between $0.01$ and $0.026$.
Thus, R--C and MD--C show closer near optimality
than R--KL and MD--KL.
 \setlength{\abovecaptionskip}{-1ex}
\begin{table}\centering
\caption{Statistics for $r_{\mathrm{R-KL}}=f^{\mathrm{R-KL}}/f^{\mathrm{OPT-KL}}$ and $r_{\mathrm{MD-KL}}=f^{\mathrm{MD-KL}}/f^{\mathrm{OPT-KL}}$}
\ra{1.2}
\begin{tabular}{@{}rrrrcrrrcrrrcrrr@{}}\toprule
& \multicolumn{3}{c}{$p=3$} & \phantom{abc}& \multicolumn{3}{c}{$p=4$} &
\phantom{abc} & \multicolumn{3}{c}{$p=5$}\\
\cmidrule{2-4} \cmidrule{6-8} \cmidrule{10-12} \cmidrule{14-16}
& $\max$ & $\mathrm{avg}$ & $\min$  && $\max$ & $\mathrm{avg}$ & $\min$ && $\max$ & $\mathrm{avg}$ & $\min$\\
\hline
\vspace*{-.4cm}\\
$r_{\mathrm{R-KL}}$\\
$n=20$ &1.000 & 0.990  & 0.789  && 1.000 &0.985  & 0.688  && 1.000 & 0.977  & 0.672 \\
$n=30$ &1.000 & 0.989  & 0.830  && 1.000 &0.988  & 0.826  && 1.000 & 0.983  & 0.795 \\
$n=40$ &1.000 & 0.989  & 0.729  && 1.000 & 0.985  & 0.842  && 1.000 & 0.983  & 0.817 \\
\vspace*{-.4cm}\\
$r_{\mathrm{MD-KL}}$\\
$n=20$&  1.000 & 0.992  & 0.744  && 1.000 & 0.982  & 0.688  &&  1.000 & 0.975  & 0.672\\
$n=30$& 1.000 & 0.989  & 0.809  && 1.000 & 0.987  & 0.832  && 1.000 &  0.981  & 0.742 \\
$n=40$& 1.000 & 0.985  & 0.729  && 1.000 & 0.980  & 0.802  && 1.000 & 0.981  & 0.834 \\
\bottomrule
\end{tabular}
\end{table}
 \setlength{\belowcaptionskip}{-1ex}
\setlength{\belowcaptionskip}{-1ex}
\begin{table}\centering
\caption{Statistics for $r_{\mathrm{R-C}}=f^{\mathrm{R-C}}/f^{\mathrm{OPT-C}}$ and $r_{\mathrm{MD-C}}=f^{\mathrm{MD-C}}/f^{\mathrm{OPT-C}}$}
\ra{1.2}
\begin{tabular}{@{}rrrrcrrrcrrrcrrr@{}}\toprule
& \multicolumn{3}{c}{$p=3$} & \phantom{abc}& \multicolumn{3}{c}{$p=4$} &
\phantom{abc} & \multicolumn{3}{c}{$p=5$}\\
\cmidrule{2-4} \cmidrule{6-8} \cmidrule{10-12} \cmidrule{14-16}
& $\max$ & $\mathrm{avg}$ & $\min$  && $\max$ & $\mathrm{avg}$ & $\min$  && $\max$ & $\mathrm{avg}$ & $\min$ & \\
\hline
\vspace*{-.4cm}\\
$r_{\mathrm{R-C}}$\\
$n=20$&  1.000 & 0.994  & 0.831  &&  1.000 & 0.992  & 0.789  &&  1.000 & 0.995  & 0.850 \\
$n=30$&  1.000 & 0.997  & 0.880  &&  1.000 & 0.997  & 0.868  &&  1.000 & 0.995  & 0.883 \\
$n=40$&  1.000 & 0.998  & 0.959  &&  1.000 & 0.995  & 0.936  &&  1.000 & 0.997  & 0.959 \\
\vspace*{-.4cm}\\
$r_{\mathrm{MD-C}}$\\
$n=20$&  1.000 & 0.997  & 0.835  &&  1.000 & 0.995  & 0.874  &&  1.000 & 0.996  & 0.918 \\
$n=30$&  1.000 & 0.995  & 0.874  &&  1.000 & 0.997  & 0.892  &&  1.000 & 0.995  & 0.928 \\
$n=40$&  1.000 & 0.998  & 0.931  &&  1.000 & 0.994  & 0.933  &&  1.000 & 0.994  & 0.953 \\
\bottomrule
\end{tabular}
\end{table}
\setlength{\belowcaptionskip}{-1ex}

\mypar{Larger examples: Comparison with the best randomly generated selection}
We now consider larger $n$ and $p$, when computing the optimum by
the exhaustive search is infeasible. Similarly as in subsection~\ref{subsect-test-alg-uncert} (larger examples), we randomly generate $10^5$ sensor selections and find the maximal objective function $f^{\mathrm{BEST-RAND-KL}}$ over these selections. For a fixed
 pair $n$ and $p$,
  we generate $K_{\mathrm{exper}}=100$
   sets of the distribution parameters $m_i,\,S_i,\,i=0,1$;
    for each set of the distribution parameters,
     we evaluate $\rho_{\mathrm{R-KL}}$, $\rho_{\mathrm{MD-KL}}$, and
      $\rho_{\mathrm{R-C}}$, $\rho_{\mathrm{MD-C}}$.
       We are also interested in the ratio of the computational time
       of the R--KL (resp. MD--KL) algorithm over the computational time
       of checking $10^5$ random selections (and the analogous quantities
       for the Chernoff-based selection).

Table~{VI} shows the average (avg), the maximal (max), and the minimal (min) values
of $\rho_{\mathrm{R-KL}}$ and $\rho_{\mathrm{MD-KL}}$ over $100$ data sets $m_i, \,S_i,\,i=0,1$;
 Table~{VII} (left) shows the computational time ratios. We can see that both R--KL and
 MD--KL outperform best random selection strategy, as they achieve larger
  objective function while reducing computational time. For example,
   for $n=80$, $p=0.1 \times n=8$, R--KL can give $54\%$ larger
    objective function, while it has $3$ times smaller computational time.
    MD--KL is better than R--KL, as it
    shows comparable performance
    in terms of the objective function; at the same time,
    it significantly reduces the computational time.
    Thus, for sensor selection without uncertainties,
    MD--KL is a very good tool that can handle large problem instances. Similar conclusions
     hold for the Chernoff based counterpart algorithms (see Table~{VII} and Table~{VIII} (right)).
     Chernoff based algorithms have larger computational time
      than Kullback-Leibler based algorithms, which is expected due
      to additional maximization over variable~$s$ (see \eqref{op-C}). Section~{V} shows
       generally that Chernoff criterion has some advantage over Kullback-Leibler criterion,
        which trades off the computational requirements to solve for these criteria.
 \setlength{\belowcaptionskip}{-1ex}
\begin{table}\centering
\caption{Statistics for $\rho_{\mathrm{R-KL}}=f^{\mathrm{R-KL}}/f^{\mathrm{BEST-RAND-KL}}$ and $\rho_{\mathrm{MD-KL}}=f^{\mathrm{MD-KL}}/f^{\mathrm{BEST-RAND-KL}}$}
\ra{1.2}
\begin{tabular}{@{}rrrrcrrrcrrr@{}}\toprule
& \multicolumn{3}{c}{$p=0.1 \times n$} & \phantom{abc}& \multicolumn{3}{c}{$p=0.2 \times n$} &
\phantom{abc} & \multicolumn{3}{c}{$p=0.3 \times n$}\\
\cmidrule{2-4} \cmidrule{6-8} \cmidrule{10-12}
 & $\mathrm{avg}$ & $\max$ & $\min$ && $\mathrm{avg}$ & $\max$ & $\min$ && $\mathrm{avg}$ & $\max$ & $\min$\\
\hline
\vspace*{-.12cm}\\
$\rho_{\mathrm{R-KL}}$\\
$n=50$& 1.069 & 1.246 & 0.855 &&  1.225 & 1.577 & 1.072 && 1.274 & 1.733 & 1.071\\
$n=80$& 1.298 & 1.544 & 1.162 && 1.474 & 1.860 & 1.233 && 1.472 & 1.962 & 1.229\\
$n=100$& 1.427 & 1.705 & 1.169 && 1.556 & 1.739 & 1.337 && 1.598 & 1.894 & 1.378\\
\vspace*{-.12cm}\\
$\rho_{\mathrm{MD-KL}}$\\
$n=50$& 1.072 & 1.246 & 0.919 && 1.223 & 1.464 & 1.002 && 1.265 & 1.684 & 1.077\\
$n=80$& 1.299 & 1.544 & 1.130 && 1.475 & 1.782 & 1.248 && 1.468 & 2.002 & 1.225\\
$n=100$& 1.429 & 1.705 & 1.169 && 1.563 & 1.864 & 1.384 && 1.617 & 1.973 & 1.374\\
\bottomrule
\end{tabular}
\end{table}
 \setlength{\belowcaptionskip}{-1ex}
 \setlength{\belowcaptionskip}{-1ex}
\begin{table}\centering
\caption{Statistics for $\rho_{\mathrm{R-C}}=f^{\mathrm{R-C}}/f^{\mathrm{BEST-RAND-C}}$ and $\rho_{\mathrm{MD-C}}=f^{\mathrm{MD-C}}/f^{\mathrm{BEST-RAND-C}}$}
\ra{1.2}
\begin{tabular}{@{}rrrrcrrrcrrr@{}}\toprule
& \multicolumn{3}{c}{$p=0.1 \times n$} & \phantom{abc}& \multicolumn{3}{c}{$p=0.2 \times n$} &
\phantom{abc} & \multicolumn{3}{c}{$p=0.3 \times n$}\\
\cmidrule{2-4} \cmidrule{6-8} \cmidrule{10-12}
& $\mathrm{avg}$ & $\max$ & $\min$ && $\mathrm{avg}$ & $\max$ & $\min$ && $\mathrm{avg}$ & $\max$ & $\min$\\
\hline
\vspace*{-.12cm}\\
$\rho_{\mathrm{R-C}}$\\
$n=50$&   1.074 & 1.214 & 0.992 && 1.182 & 1.341 & 1.072 &&  1.194 & 1.307 & 1.095\\
$n=80$&   1.262 & 1.454 & 1.132 && 1.357 & 1.621 & 1.199 &&  1.338 & 1.464 & 1.206\\
$n=100$&  1.375 & 1.517 & 1.251 && 1.447 & 1.616 & 1.299 &&  1.402 & 1.629 & 1.287\\
\vspace*{-.12cm}\\
$\rho_{\mathrm{MD-C}}$\\
$n=50$&  1.074  & 1.214 & 0.991 && 1.182 & 1.341 & 1.038 && 1.195 & 1.307 & 1.100\\
$n=80$&  1.262  & 1.471 & 1.132 && 1.357 & 1.621 & 1.194 && 1.338 & 1.464 & 1.212\\
$n=100$& 1.375  & 1.517 & 1.254 && 1.445 & 1.616 & 1.296 && 1.403 & 1.632 & 1.296\\
\bottomrule
\end{tabular}
\end{table}
 \setlength{\belowcaptionskip}{-1ex}
\setlength{\abovecaptionskip}{-1ex}
\begin{table}\centering
\caption{Average time ratios,  $K_{\mathrm{exper}}$=100; Left: KL algorithms; Right: C algorithms}
\ra{1.2}
\begin{minipage}{3.2in}
\begin{tabular}{@{}rrrr@{}}\toprule
 & $p=10\%$ & $p=20\%$ & $p=30\%$ \\
\midrule
\vspace*{-.12cm}\\
R--KL\\
$n=50$&  0.070 &  0.104 & 0.107 \\
$n=80$&  0.276 & 0.295 & 0.291 \\
$n=100$&  0.417 & 0.472 & 0.349\\
\vspace*{-.12cm}\\
MD--KL\\
$n=50$&  0.003 & 0.005 & 0.006 \\
$n=80$&  0.007 & 0.012 & 0.014 \\
$n=100$&  0.011 & 0.017 & 0.021\\
\bottomrule
\end{tabular}
\end{minipage}
\begin{minipage}{3.2in}
\ra{1.2}
\begin{tabular}{@{}rrrr@{}}\toprule
 & $p=10\%$ & $p=20\%$ & $p=30\%$ \\
\midrule
\vspace*{-.12cm}\\
R--C\\
$n=50$&  0.105 & 0.146 &  0.153 \\
$n=80$&  0.472 & 0.549 & 0.507 \\
$n=100$& 0.875 & 0.990  & 0.721\\
\vspace*{-.12cm}\\
MD--C\\
$n=50$&   0.002 & 0.004 & 0.006\\
$n=80$&  0.006 & 0.011 & 0.014\\
$n=100$&  0.010 & 0.017 & 0.021\\
\bottomrule
\end{tabular}
\end{minipage}
\end{table}
 \setlength{\belowcaptionskip}{-1ex}

\vspace{-1em}
\section{Conclusions}
In this paper, we addressed the problem of finding the most informative subset of $p$ out of $n$ sensors, for the task of deciding between the two possible hypothesis on the monitored environment. We proposed two different information theoretic criteria for the best sensor selection: the Kullback-Leibler distance and the Chernoff distance between the distributions induced by selected sensors. We tackled the case where the distributions are Gaussian, but the mean vectors are known only up to confidence regions. We formulated the corresponding maxmin optimization problems, and developed the R--KL and R--C algorithms, that efficiently solve the problems with complexity $\O(n^3p+np^4)$. We also addressed the case when the mean vectors are known and, for this case, we exploited the structure of the problems to develop more efficient algorithms, MD--KL and MD--C, of complexity $\O(n^3+np^3)$. We performed Monte-Carlo based experiments to test both the proposed sensor selection criteria and the sensor selection algorithms. Numerical studies of the criteria show that the Kullback-Leibler based and the Chernoff based selections have near optimal performance, both in the Neyman-Pearson and Bayes sense. The performance of our algorithms we compared with 1) the optimal sensor selections, when the exhaustive search to compute them is feasible (smaller $n$ and $p$); and 2) with best random selections, when $n$ and $p$ are large. Comparison with the exhaustive search shows that proposed algorithms in many cases find the optimal selection and, on average, are at most $5\%$ below the optimal value; at the same time, computational savings are significant. For larger problems, simulation results demonstrate that our algorithm outperforms random searches, once an upper bound on computational time is set.
\vspace{-1em}
\appendix
\section{Appendix}
\label{Appendix}
\vspace{-0.5em}
\subsection{Proof of NP hardness of optimization problems~\eqref{op-KL} and~\eqref{op-C}}
\label{App-NP-hardness}
We will prove that both problems~\eqref{op-KL} and~\eqref{op-C}
are NP hard by reducing them to the maximal clique problem (MQP), which is known to be NP hard, see, e.g.,~\cite{NP-book}.
We first define the MQP. Consider an undirected, simple (i.e., without self-loops) graph $\mathcal{G} = \left( \mathcal{N}, \mathcal{E}  \right)$,
 where $\mathcal{N}$ is the set of vertices with cardinality $|\mathcal{N}|=n$, and $\mathcal{E}$ is the set of
  undirected edges $\{i,j\}$, $|\mathcal{E}|=m$. A clique of the graph $\mathcal{G}$, of size $p$, is a complete subgraph of
  $\mathcal{G}$ that has $p$ vertices. The decision version of the maximum clique problem is as follows:

   \textbf{MQP: The maximal clique problem:} ``For given graph $\mathcal{G}$ and a positive integer $p$, $1\leq  p\leq n$, determine whether
   $\mathcal{G}$ has a clique of size at least $p$.''

  We conduct a reduction to MQP by attaching to a graph $\mathcal{G}$
  a $n \times n$ matrix $S\left( \mathcal{G}\right)$ of a special structure. Namely, we define a positive definite matrix $S\left(\mathcal{G}\right)$ as follows:
  \begin{equation}
  \label{eqn-S-matrix}
  \left[S\left( \mathcal{G} \right) \right]_{ij}=
  \left\{ \begin{array}{rl}
 2n &\mbox{ ;\,if $i=j$} \\
 -1 &\mbox{;\,if $i\neq j$, $\{i,j\} \in \mathcal{E}$}\\
  0 &\mbox{;\, otherwise}
       \end{array}. \right.
  \end{equation}
The matrix $S\left(  \mathcal{G} \right)$ is positive definite
 because it has positive diagonal elements and it is strictly diagonally dominant. Now, fix an integer $p$, $1 \leq p\leq n$, and consider a set of matrices $\mathcal{A}_p$ defined as $\mathcal{A}_p=
   \left\{ A \in {\mathbb R}^{p \times p}:\,\,A=A^\top,\,\,A_{ii}=2n,\,\,\forall i,\,\,
   A_{ij} \in \{0,-1\},\,\,i \neq j   \right\}$. Clearly, all matrices in
   $\mathcal{A}_p$ are positive definite, as they are strictly diagonally dominant,
   with positive diagonal entries.
   Denote with ${\bfone}_p$ the column vector with all entries equal to $1$ and define the function $\textbf{s.e.i.}:\,{\mathcal{A}_p} \rightarrow \mathbb R$
    as $\textbf{s.e.i.}(A)={\bfone}_p^\top A^{-1} {\bfone}_p$ ($\textbf{s.e.i.}$ is the sum of the elements in the inverse.) We have the following result on the matrices
   in $\mathcal{A}_p$.
\vspace{-0.7em}
   \begin{lemma}
   \label{lemma-S-matrices}
   For all matrices $A \in \mathcal{A}_p$, there holds:
   \begin{enumerate}
   \item
   Denote $B:=A^{-1}$. Then, for all $i,j$ $B_{ij} \geq 0$.
    \item
    If $A_{ij}=0$, then $\textbf{s.e.i.}\left( A - h_i h_j^\top - h_j h_i^\top \right)  \geq \textbf{s.e.i.}(A) .$ (Recall that $h_i$ denotes $i$th canonical vector.)
    \item
    $\textbf{s.e.i.}(A) \leq \frac{p}{2n-p+1},$
    where the equality holds if and only if $A = A^\star := 2n I - 1 1^\top + I$.
   \end{enumerate}
   \end{lemma}

   \begin{proof}
   The claim~1) in Lemma~\ref{lemma-S-matrices} follows from the fact that the matrix $A$ is an $M$ matrix~\cite{Diagonally-Dominant},
   and, consequently, the matrix $B=A^{-1}$ has all
   entries greater than or equal to zero (e.g.,~\cite{Diagonally-Dominant}).
We now show the claim 2) in Lemma~\ref{lemma-S-matrices}. Remark first that the function $\textbf{s.e.i.}(\cdot)$ is convex and differentiable on the set of positive definite matrices. Applying the first order Taylor expansion lower bound at $A$, we get:
\begin{equation}
\label{eq-first-order-Taylor-LB}
\textbf{s.e.i.}\left( A - h_i h_j^\top - h_j h_i^\top \right)\geq \textbf{s.e.i.}(A)+\mathrm{tr}\left(\bold\nabla \textbf{s.e.i.}(A)\left(- h_i h_j^\top - h_j h_i^\top\right) \right),
\end{equation}
where $\bold\nabla \textbf{s.e.i.}(A)$ stands for the (matrix form) gradient of $\textbf{s.e.i.}(\cdot)$ at $A$ and is equal to $-A^{-1}{\bfone}_p {\bfone}_p^{\top} A^{-1}$. Now, by claim~1), all entries of $A^{-1}$ are nonnegative, and, thus, the second term on the right hand side of the inequality~\eqref{eq-first-order-Taylor-LB} is nonnegative as well. This completes the proof of claim~$2$.

We proceed and prove the claim 3) in Lemma~\ref{lemma-S-matrices}. From claim~2) we know that the more $-1$'s a matrix $A\in\mathcal{A}_p$ has on its off-diagonal entries, the higher the value $\textbf{s.e.i.}(A)$ can be. Therefore, $A^\star$ is a maximizer of $\textbf{s.e.i.}$ over the set $\mathcal{A}_p$. Also, it is straightforward
to check that $\textbf{s.e.i.}(A^\star)=\frac{p}{2n-p+1}$. We will show next that $A^\star$ is in fact the only maximizer of $\textbf{s.e.i.}$ (over $\mathcal{A}_p$). To show this, it suffices to show that, for any choice of $1\leq i,j\leq p, i\neq j$, the following strict inequality holds:
\begin{equation}
\label{eq-A-star-the-maximizer}
\textbf{s.e.i.} \left(A^{\star} + h_i h_j^\top + h_j h_i^\top \right)<\textbf{s.e.i.} (A^{\star}).
\end{equation}
To this end represent the matrix $h_i h_j^\top + h_j h_i^\top$ as
         $ h_i h_j^\top + h_j h_i^\top =  H C H^\top$,
where $H=\left[h_i\,h_j\right]\in{\mathbb R}^{n\times 2}$ and
$C\in{\mathbb R}^{2\times 2},\,C_{12}=C_{12}=1,\,C_{11}=C_{22}=0.$
Using the matrix inversion lemma,
 we get:
 \begin{eqnarray}
 \label{eq-A-star-inv-lemma}
 \textbf{s.e.i.} \left(A^\star + h_i h_j^\top + h_j h_i^\top \right) &=& {\bfone}_p^\top B^\star {\bfone}_p - {\bfone}_p B^\star H \left(  C^{-1} + H^\top B^\star H \right)^{-1} H^\top B^\star {\bfone}_p.
 \end{eqnarray}
After straightforward algebra, we obtain:
$${\bfone}_p B^\star H \left(  C^{-1} + H^\top B^\star H \right) H^\top B^\star {\bfone}_p=\frac{2n+1}{n+1+\frac{1}{2n-p+1}}.$$
%
%
Since this term is greater than zero for all $1\leq p\leq n$, the inequality~\eqref{eq-A-star-the-maximizer} follows. This completes the proof of claim~3) and the proof of the Lemma.
\end{proof}

We proceed with the proof of Theorem~($1$). The decision version of~\eqref{op-KL},
 for $k_1=k_0=+\infty$, is:

  \textbf{D--KL: Decision version of~~\eqref{op-KL}} ``For given data: 1) vectors $m_0, m_1 \in {\mathbb R}^n$;
   2) positive definite matrices $S_0$ and $S_1$; 3) positive integer $p$, $p \leq n$;
    and 4) a number $f_{\mathrm{KL}}^{\bullet}$, determine whether there is a $n \times p$
     sensor selection matrix $E$, such that $f_{\mathrm{KL}}(E)$ defined in eqn.~\eqref{KL-cost-fcn}
      is at least $f_{\mathrm{KL}}^{\bullet}$.''

 We now reduce the MQP to D-KL. Consider a simple, undirected graph $\mathcal{G}$ and consider MQP of determining whether there is a clique in $\mathcal{G}$ of the size at least $p$. Define the matrix $S(\mathcal{G})$
   as in eqn.~\eqref{eqn-S-matrix}. Consider an instance of D--KL, for some fixed $p$, with the following
   data: 1) $m_1={\bfone}_n$, $m_0={\bfzero}_n$; 2) $S_1=S_0=S(\mathcal{G})$; 3) $p$; and 4)
   $f_{\mathrm{KL}}^{\bullet}=\frac{1}{2} \frac{p}{2n-p+1}$. Now, the answer
   to D--KL is YES (resp. NO) if and only if
    $\mathcal{G}$ has (resp does not have) a clique of size at least $p$. Thus, MQP problem
     is reduced to D--KL.

MQP problem reduces to the decision version of problem~\eqref{op-C} (denoted by D--C), for $k_1=k_0=+\infty$, in a very similar way as it reduces to D--KL, by considering the instance of D--C with the data $m_1,m_0,S_1,S_0,p,$ same as we considered for D--KL, but $f_{\mathrm{KL}}^{\bullet}=\frac{1}{2} \frac{p}{2n-p+1}$ is replaced by $f_{\mathrm{C}}^{\bullet}=\frac{1}{8} \frac{p}{2n-p+1}.$
\vspace{-2em}
\subsection{Proof that the Kullback-Leibler and Chernoff distances are not submodular functions}
\label{App-Not-submodular-CounterEx}
See~\cite{Guestrin} for the definition of a submodular function. Consider two Gaussian distributions ${\mathcal N}(m_i,S_i)$, $i=0,1$, with parameters $m_0=m_1$, $S_0=I_3$ and $S_1=I_3+\epsilon(h_2h_3^{\top}+h_3h_2^{\top})$ ($h_2$, $h_3\in{\mathbb{R}}^3$), where $0<\epsilon <1$.
Let $E_1=h_1\,\in{\mathbb{R}}^3$, $E_{13}=\left[h_1\,h_3\right]$, $E_{12}=\left[h_1\,h_2\right]$ and $E_{123}=I_3$.

Computing the Kullback Leibler distance for selections $E_1$, $E_{12}$ and $E_{13}$ we get $f_{\mathrm{KL}}(E_1)=f_{\mathrm{KL}}(E_{12})=f_{\mathrm{KL}}(E_{13})=0$, whereas $f_{\mathrm{KL}}(E_{123})=-\frac{1}{2}\log (1-{\epsilon}^2) > 0$.
Thus, $f_{\mathrm{KL}}(E_{13})-f_{\mathrm{KL}}(E_{1})
<f_{\mathrm{KL}}(E_{123})-f_{\mathrm{KL}}(E_{12})$ which proves that function $f_{\mathrm{KL}}$ is not submodular.
The proof for the case of Chernoff distance can be done in similar way.
\vspace{-1em}
\subsection{Solution to problem~\eqref{op-KL-no-uncert-relaxation-reformulated}}
\label{App-Equal-Cov-KL}
As explained in subsection~\ref{subsubsec-KL-eq-means},
the cost function can be written as $\frac{1}{2}\sum_{i=1}^{p}\phi_{\mathrm{KL}} \left(\lambda_i\left(P^{\top} SP\right)\right)$, where $\phi_{\mathrm{KL}}(x)=x-\log x-1$. Now, invoking Poincar\'{e} separation theorem, we can write~\eqref{op-KL-no-uncert-relaxation-reformulated} as:
\begin{equation}
\begin{array}[+]{ll}
\mbox{maximize} & \frac{1}{2}\sum_{i=1}^{p}\phi_{\mathrm{KL}}(x_i)\\
\mbox{subject to} &  \lambda_i(S)\leq x_i \leq \lambda_{n-p+i}(S),\;i=1,...,p\\
& x_1 \leq x_2 \leq \ldots \leq x_p \\
& \exists P\in \mathrm{R}^{n\times p} \mbox{\;s.t\;}. P^{\top} P = I_{p} \mbox{\;and\;} x_i=\lambda_i(P^{\top} SP) \mbox{\;for\;} i=1,...,p.
\end{array}\label{opt-KL-equal-means-rewrit}
\end{equation}
Next, we relax the problem~\eqref{opt-KL-equal-means-rewrit} in the sense that we do not require that variables $x_i$ are generated via the Stiefel matrix $P$:
\begin{equation}
\begin{array}[+]{ll}
\mbox{maximize} & \frac{1}{2}\sum_{i=1}^{p}\phi_{\mathrm{KL}}(x_i)\\
\mbox{subject to} &  \lambda_i(S)\leq x_i \leq \lambda_{n-p+i}(S),\;i=1,...,p\\
& x_1 \leq x_2 \leq \ldots \leq x_p. \\
\end{array}\label{opt-KL-equal-means-relax}
\end{equation}
We now focus on problem~\eqref{opt-KL-equal-means-relax}. Function $\phi_{\mathrm{KL}}(\cdot)$ is convex, with minimum equal to zero attained at point $1$. Thus, its maximum over a compact interval is always at a boundary point. Also, if both boundary points are greater (resp. smaller) than $1$, then the maximizer is the right (resp. left) boundary point. Therefore, if $\lambda_i(S)\geq 1$, then optimal $x_i$ equals $\lambda_{n-p+i}(S)\geq 1$. Similarly, if $\lambda_{n-p+i}(S)\leq 1$, optimal $x_i$ equals $\lambda_i(S)$. If we rule out all eigenvalue intervals that fall in either of the two previous categories, we are left with intervals where $1$ is in their interiors. We now focus on those eigenvalue intervals, i.e. on those $i=1,\ldots,p$ s.t. $\lambda_i(S)<1$ and $\lambda_{n-p+i}(S)> 1$. It can be shown that there will exist a switching index $i_{\mathrm{switch}}\in \{0,1,\ldots,p\}$  such that for all indices on the left of (or equal to) $i_{\mathrm{switch}}$, optimal $x_i$ is on the left boundary point, while for all indices on the right of $i_{\mathrm{switch}}$ optimal $x_i$ is on the right boundary point. Summing up, a solution $x^{\star}$ of~\eqref{opt-KL-equal-means-relax} will be always of the form $x^{\star} =(\lambda_1(S),\, \ldots,\,
\lambda_{i_{\mathrm{switch}}}(S),\, \lambda_{n-p+i_{\mathrm{switch}}+1}(S),\, \ldots,\, \lambda_n(S))^\top$. (Case $i_{\mathrm{switch}}=0$ corresponds to $x^\star = (\lambda_{n-p+1}(S),\, \ldots,\, \lambda_n(S))^\top$.)

We now construct solution of~\eqref{opt-KL-equal-means-rewrit} from $x^\star$--solution of~\eqref{opt-KL-equal-means-relax}; $x^\star$ consists of a subset of eigenvalues of the matrix $S$; moreover, it is given by $x^{\star} =(\lambda_1(S),\, \ldots,\, \lambda_{i_{\mathrm{switch}}}(S),\, \lambda_{n-p+i_{\mathrm{switch}}+1}(S),\, \ldots,\, \lambda_n(S))^\top$,
for some $i_{\mathrm{switch}} \in \{0,1,...,p\}$ (that must be determined). Therefore, there exists a matrix $P^\star$ that generates $x^{\star}$, and hence solves~\eqref{opt-KL-equal-means-rewrit}. The columns of the matrix $P^\star$ are simply orthonormal eigenvectors of $S$ corresponding to the eigenvalues $\lambda_1(S),\, \ldots,\, \lambda_{i_{\mathrm{switch}}}(S),\, \lambda_{n-p+i_{\mathrm{switch}}+1}(S),\, \ldots,\, \lambda_n(S)$.

We remark that the solution $P^{\star}$ of~\eqref{op-KL-no-uncert-relaxation-reformulated} is obtained in at most $p+1$ steps.
 The number of steps is smaller when some of the interlacing eigenvalue intervals do not contain
  $1$. If the number of such intervals is $k$, then the number of steps that our algorithm requires is exactly $p-k+1$.

\nopagebreak[4]
\bibliographystyle{IEEEtran}
\bibliography{IEEEabrv,Bibliography}
\end{document}